\documentclass[%
 reprint,
 pra,
 floatfix,
 amsmath,amssymb,
 aps
]{revtex4-2}

\usepackage{xcolor}
\usepackage{hyperref}
\usepackage{graphicx}
\usepackage{dcolumn}
\usepackage{bm}
\usepackage{physics}
\usepackage{xspace}

\newcommand{\opsigma}{\hat{\sigma}}
\newcommand{\Fe}{\ensuremath{{}^{57}\mathrm{Fe}}\xspace}

\newcommand{\Sn}{\ensuremath{{}^{119}\mathrm{Sn}}\xspace}
\newcommand{\Sc}{\ensuremath{{}^{45}\mathrm{Sc}}\xspace}
\newcommand{\SR}{\ensuremath{\Gamma_\mathrm{SR}}\xspace}
\newcommand{\CLS}{\ensuremath{\Delta_\mathrm{CLS}}\xspace}

\renewcommand{\vec}[1]{\ensuremath{{\bm{#1}}}}

\allowdisplaybreaks[2]

\begin{document}
\title{Inverse design of artificial two-level systems with M\"ossbauer nuclei in thin-film cavities}

\author{Oliver Diekmann}
\email{oliver.diekmann@mpi-hd.mpg.de}

\author{Dominik Lentrodt}

\author{J\"org Evers}
\email{joerg.evers@mpi-hd.mpg.de}

\affiliation{%
 Max-Planck-Institut f\"ur Kernphysik, Saupfercheckweg 1, 69117 Heidelberg, Germany
}%

\date{\today}

\begin{abstract}
Thin-film cavities containing layers of Mössbauer nuclei have been demonstrated to be a rich platform for x-ray quantum optics. At low excitation, these systems can be described by effective few-level schemes, thereby providing  tunable artificial quantum systems at hard x-ray energies. With the recent advent of an ab initio theory, a numerically efficient description of these systems is now possible. On this basis, we introduce the inverse design and develop a comprehensive optimization for an archetype system with a single resonant layer, corresponding to an artificial two-level scheme. We discover a number of qualitative insights into x-ray photonic environments for nuclei that  will likely impact the design of future x-ray cavities and thereby improve their performance. The presented methods readily generalize beyond the two-level case and thus provide a clear perspective towards the inverse design of more advanced tunable x-ray quantum optical level schemes.
\end{abstract}

\maketitle

\section{Introduction}\label{sec:Introduction}
The conventional approach to describe physical systems is to define the system, and to subsequently derive its functionality. In most relevant cases, the functionality depends on the system's design in a nontrivial way, such that the development of novel or improved functionality is challenging and often founds on intuition. Opposed to that, specifying the desired properties at the outset of the procedure, and trying to afterwards find a suitable  apparatus, is known as \textit{inverse design}. Exploring the yet unknown, inverse design is not always trivially realized as a corresponding setup might not exist. Also in the realm of nanophotonics (see~\cite{molesky_inverse_2018,https://doi.org/10.1002/lpor.201000014} and references therein) and light-matter interactions~\cite{bennett_inverse_2020}, inverse design has become a paradigm sought after. 

\begin{figure*}[t]
    \includegraphics[width=0.9\textwidth]{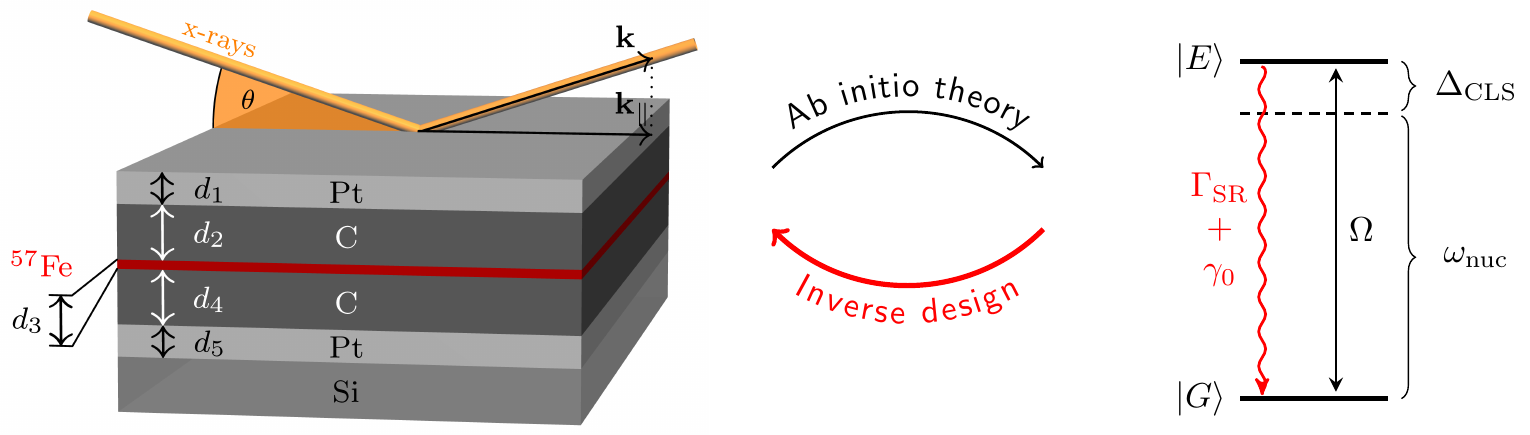}
    \caption{(Color online) Schematic setup of the archetypal thin-film cavity system and illustration of its relation to the artificial two-level scheme. The cavity consists of a stack of different layers (thicknesses and exemplary materials are indicated) and a single, thin layer of M\"ossbauer nuclei (here: \Fe). The cavity is illuminated in grazing incidence with x-rays near-resonant to the nuclear transition frequency. $\theta$, $\vec{k}$  and $\vec{k}_\parallel$ are the x-ray incidence angle, the wave vector, and its projection onto the cavity surface, respectively. At low probing intensities, the system features an effective description as an artificial two-level system. As compared to the bare two-level system of individual nuclei with transition frequency $\omega_\mathrm{nuc}$ and decay rate $\gamma_0$, it is driven by a cavity- and collectively enhanced Rabi-frequency $\Omega$, and features a level shift \CLS and a decay rate enhancement \SR. A recently developed ab initio theory allows one to readily derive the artificial level scheme from a given cavity structure~\cite{lentrodt_ab_2020}. In this work, we introduce and develop the \textit{inverse design} of artificial two-level systems, i.e., we determine cavity structures suitable to the realization of desired two-level schemes.}
    \label{fig:schematicSetup}
\end{figure*}

In this work, we introduce and develop the inverse design of artificial x-ray quantum level schemes which are realized with M\"ossbauer nuclei in thin-film cavities probed in grazing incidence, see Fig. \ref{fig:schematicSetup}. These systems constitute an intriguing platform for quantum optics in the x-ray regime \cite{Adams2012,adams_x-ray_2013,kuznetsova_quantum_2017,adams_scientific_2019,yoshida_quantum_2021}.
We focus the design on the nm-scale layer thicknesses and materials as well as the x-ray incidence angle, rather than allowing for photonic structures with arbitrary shapes. This is motivated by the present state-of-the-art in cavity fabrication, but also by the fact that hard x-ray wavelengths considered here are on or below the scale of the lattice parameters of corresponding materials---for example, the resonance wavelength of the M\"ossbauer transition in ${}^{57}$Fe is 86~pm, whereas the lattice constant of $\alpha$-Fe is 287~pm.

Resonances in M\"ossbauer nuclei are distinct from corresponding electronic x-ray resonances by their exceptionally narrow line widths, owing to the recoilless absorption and emission of photons \cite{mossbauer_kernresonanzabsorption_1959,Hannon1999,moessbauer_story}. This has the advantage of  desirably long coherence lifetimes of the nuclear excitations, but on the other hand severely limits the possibility to strongly drive or control nuclear x-ray transitions even at modern x-ray sources. As a result, it remains challenging to implement advanced nuclear multi-level schemes featuring several transitions driven by probe or control fields.  One of the most successful approaches to overcome this restriction so far is to influence the nuclear properties and dynamics with tailor-made photonic environments such as x-ray cavities~\cite{yoshida_quantum_2021}. Note that complementary approaches towards nuclear quantum optics also have  made substantial progress, see~\cite{Vagizov1990,PhysRevB.47.7840,PhysRevB.52.10268,Shvydko1996,Odeurs2002,Vagizov2014,Liao2016,Heeg2017,Sakshath2017,Chumakov2018,Heeg2021,Bocklageeabc3991,chen_transient_2021} and references therein.

The key idea behind the x-ray cavity approach is that in the experimentally relevant low-excitation regime, the joint nuclei-cavity system can be shown to feature an effective description as a tunable x-ray quantum optical few-level scheme \cite{rohlsberger_collective_2010,rohlsberger_electromagnetically_2012,heeg_x-ray_2013, heeg_collective_2015, lentrodt_ab_2020,kong_greens-function_2020}. By choosing a suitable cavity setup, the structure  of the few-level scheme, the transition frequencies and the decay rates can be controlled and changed as compared to the bare nuclear properties. Furthermore, the cavity environment can induce coherent couplings between the different states of the few-level system, thus compensating for the lack of direct x-ray control fields implemented with external sources.  Experimentally, a broad range of quantum optically inspired setups has been realized using this approach. Examples include the collective Lamb shift (CLS) and superradiance (SR) \cite{rohlsberger_collective_2010} (which have been intensely studied in other systems~\cite{friedberg_frequency_1973,PhysRev.120.513,Kagan1979,Hannon1999,scully_collective_2009, ruostekoski_emergence_2016, keaveney_cooperative_2012, guerin_light_2017, peyrot_collective_2018, roof_observation_2016, bromley_collective_2016}),  electromagnetically induced transparency \cite{rohlsberger_electromagnetically_2012}  (see~\cite{RevModPhys.77.633} for a review), vacuum generated coherences \cite{heeg_vacuum-assisted_2013}~(see~\cite{Ficek2005,Kiffner2010} for reviews),  the realization of strong coupling \cite{haber_collective_2016}, Rabi-oscillations \cite{haber_rabi_2017} and sub-luminal propagation of x-ray pulses \cite{heeg_tunable_2015} (see, e.g.,~\cite{Boyd1074,Khurgin2018}).
Beyond that, a number of schemes involving x-ray cavities have been theoretically proposed (see, e.g.,~\cite{Joshi2015,heeg2016inducing,Kong2016,Huang2017} and references therein), as well as related schemes in other photonic environments~\cite{PhysRevLett.122.123608,vassholz_observation_2021,PhysRevLett.126.053201,D1SC01774H,PhysRevResearch.3.033063,chen_transient_2021}.

Over the past decade, the quantum optical description of the cavity setups in terms of few-level systems has been continuously advanced~\cite{heeg_x-ray_2013, heeg_collective_2015,kong_greens-function_2020,lentrodt_ab_2020}.
A decisive step towards the inverse design of artificial x-ray few-level systems was recently taken with the advent of an ab initio quantum optical theory~\cite{lentrodt_ab_2020}. The latter is formulated in terms of the classical electromagnetic Green's function which is known analytically~\cite{tomas_green_1995} and thus allows for the numerically efficient calculation of the effective level schemes. 

Here, we use this ab initio method  to introduce and develop the inverse design of artificial nuclear few-level schemes. We focus on two-level systems, which are realized in an archetype cavity setup featuring a single resonant nuclear layer, see Fig.~\ref{fig:schematicSetup}. As compared to the bare nuclei, the cavity, and the collective effects it mediates, may lead to a modified transition frequency and spontaneous decay rate, as well as to an enhancement of the external driving of the two-level system. 

In a first step, we determine the complete parameter space of the realizable artificial two-level systems. This already allows for inverse design, since we can attribute specific cavity geometries to any of the possible  quantum optical parameter combinations. However, the optimization towards different design goals does not always lead to experimentally relevant cavities. Therefore, in a second step, we include the visibility of the nuclear resonance in the experimentally accessible reflectance spectrum as an additional design goal in the procedure.

Performing the inverse design, and also optimizing it in terms of different layer materials and resonant isotopes, we are led to a number of qualitative and somewhat unexpected insights into x-ray cavity QED with M\"ossbauer nuclei. First, the accessible parameter space shows interesting features that can be associated with the mode structure of the cavity. Second, we find that cavities without the topmost (mirror) layer may outperform their counterparts with a mirror in relevant settings. Third, our analysis reveals that cavities featuring maximum superradiance have entirely different geometries than those with maximum field enhancement at the nuclei. By contrast, these two optimization goals coincide, e.g., in standard optical Fabry-P\'erot cavities. A closer inspection reveals that the grazing incidence operation of the x-ray cavities is responsible for this qualitative difference. Fourth, our results on the optimization of cladding and guiding layer materials suggest an increase in the tuning capabilities, but also the cavity performance more generally, upon revision of the \textit{high-$Z$ cladding / low-$Z$ guiding material} design paradigm, implicit in most cavity designs employed so far. Here, $Z$ is the atomic number of the material, which directly influences the index of refraction at x-ray energies. Instead, we find that the absorption of the cladding layer is the dominant limiting factor on the performance. Finally, by analyzing the impact of the  nuclear isotope on the artificial level scheme design, we find that the nuclear properties alone are not sufficient to determine the isotope's influence on the overall performance of a thin-film cavity system. Instead, it is crucial to also consider the isotope's impact on the photonic environment for a comprehensive assessment.

The paper is organized as follows. In Sec.~\ref{sec:theory}, we define our model, derive the two-level artificial quantum system from the cavity structure, and introduce our observables. In Sec.~\ref{sec:CLS-SR-VIS}, we start with the inverse design of the artificial two-level system. In Sec.~\ref{sec:inv-vis}, we add the reflection visibility as an additional design goal, to ensure the experimental relevance of the results. In Sec.~\ref{sec:CLS-SR-EF}, we contrast the results of Sec.~\ref{sec:inv-vis} with corresponding ones including the intra-cavity field enhancement as a design goal, and compare them to the case of a Fabry-P\'erot cavity. In Secs.~\ref{sec:Materials} and~\ref{sec:Isotopes}, we explore the roles of the layer materials and the nuclear isotope on the inverse design, respectively. Finally, Sec.~\ref{sec:conclusion} summarizes and discusses the results.

\section{\label{sec:theory}The artificial x-ray two-level system}

The key concept underlying the present work is the observation that in the low-excitation limit, thin-film cavities doped with large ensembles of resonant nuclei are analytically equivalent to suitably chosen single artificial few-level systems, see Fig.~\ref{fig:schematicSetup}. Thereby, they form a platform to realize level schemes otherwise inaccessible at hard 
x-ray energies. In the following, we summarize the equations of motion  governing the nuclei in the waveguide, explain the above-mentioned equivalence to a few-level system, and discuss the relevant observables, within the framework of macroscopic QED. The latter allows one to express the Hamiltonian describing the quantized light field in the cavity and the coupling to the nuclei embedded therein in terms of the classical electromagnetic Green's function. Details on the derivation can be found in Ref.~\cite{lentrodt_ab_2020}.

\subsection{Nuclear many-body Hamiltonian in the single-particle basis}
Corresponding to the archetypal setup of Fig.~\ref{fig:schematicSetup}, we consider two-level nuclei placed in a single thin cavity layer at depth $z$. Within the Born-Markov approximation, the system's dynamics can be described via its density operator $\rho$ using a Master equation
\begin{equation}
    \dot{\rho} = -i\comm{\hat{H}}{\rho} + \mathcal{L}[\rho]\,,
    \label{eq:Master}
\end{equation}
where we employ natural units, $\hbar=c=1$. The Hamiltonian derived from macroscopic QED, given by~\cite{asenjo-garcia_atom-light_2017,buhmann_dispersion_2012}
\begin{align}
    \hat{H}=\sum\limits_{n}\frac{\omega_{\mathrm{nuc}}}{2}{\opsigma}^z_{n}-\sum\limits_{n, n'}J^{\phantom{+}}_{nn'}{\opsigma}^+_{n^{\phantom{\prime}}}{\opsigma}^-_{n'}\notag\\-\sum\limits_{n}\left[\vec{d}^*\cdot\vec{E}_{\mathrm{in}}(\vec{r}_{n}){\opsigma}^+_{n}+\mathrm{h.c.}\right]\,, \label{eq:many1}
\end{align}
is a standard many-body Hamiltonian and can be interpreted in a straightforward way. The first term describes the excitation energy of the bare nuclei enumerated by index $n$ and characterized by the Pauli operators ${\opsigma}^z_{n},{\opsigma}^\pm_{n}$. $\omega_\mathrm{nuc}$ is the nuclear transition frequency and $\vec{d}$ the nuclear transition dipole moment. The second part denotes couplings between nuclei $n$ and $n'$ with coupling constant $J_{nn'}$ mediated by the cavity environment (the dependence of the coupling constants on the Green's function will be defined in Sec.~\ref{sec:couplings} below). The final part describes the driving of the nuclei by an externally applied classical 
electric field $\vec{E}_{\mathrm{in}}(\vec{r}_{n})$ evaluated at the position of the nuclei. Note that this field in general differs from its free-space value, due to reflections and absorption in the cavity structure, as discussed in Sec.~\ref{sec:couplings} below.

Similarly, the Lindbladian assumes a standard form, 
\begin{align}
    \mathcal{L}[\rho]=&\sum\limits_{n,n'}\frac{\Gamma_{nn'}}{2}\left[{2\opsigma}^-_{n'}\rho{\opsigma}^+_{n^{\phantom{\prime}}}-\acomm{{\opsigma}^+_{n^{\phantom{\prime}}}{\opsigma}^-_{n'}}{\rho}\right]\notag\\
    & +\mathcal{L}_\mathrm{IC}[\rho]\,, \label{eq:many2}
\end{align}
where $\acomm{\mathord{\cdot}}{\mathord{\cdot}}$ is the anti-commutator. Here, $\Gamma_{nn'}$ describes spontaneous emission in the presence of the 
cavity environment for $n=n'$, and incoherent couplings between the nuclei for $n\neq n'$. The final part $\mathcal{L}_\mathrm{IC}[\rho]$ models the
 single-nucleus decay due to internal conversion.

\subsection{Nuclear few-level Hamiltonian in the spin-wave basis}
Next, we show that in the low-excitation regime, the above many-body Master equation given by Eqs.~(\ref{eq:many1},~\ref{eq:many2}) can be rewritten in terms of an effective 
two-level system by means of a suitable basis transformation.

To motivate this basis transformation, we model the synchrotron radiation 
impinging in grazing incidence onto the cavity as a classical plane-wave electromagnetic field with wavevector $\vec{k}$. 
Due to the grazing incidence geometry, nuclei at different in-plane positions will be driven with relative phase offsets determined by the projection $\vec{k}_\parallel$ of $\vec{k}$ onto the nuclear plane (cf. Fig.~\ref{fig:schematicSetup}). We take this into account by introducing spin-wave operators
\begin{align}
 {\opsigma}^\pm_{\vec{k}_\parallel} = \sum_n\,e^{\pm i\vec{k}_\parallel\cdot\vec{r}_{n\parallel}}\, \sigma_n^\pm\,,
\label{eq:SpinWaveOps}
\end{align}
which describe the excitation (deexcitation) of an excitonic spin-wave with wavevector $\vec{k}_\parallel$ in the nuclear layer due to the applied driving field, where $\vec{r}_{n\parallel}$ is the projection of $\vec{r}_n$ onto the nuclear plane.

It can be shown that in the low-excitation limit, 
and assuming translational invariance of the system along the cavity plane, the light-matter dynamics in the cavity preserves the parallel wave-vector $\vec{k}_\parallel$~\cite{lentrodt_ab_2020}. This is consistent with 
the expectation that in reflecting light on the cavity, the angles of 
incidence and reflection coincide for a translationally invariant structure under plane-wave illumination.

As a result, we can rewrite the Hamiltonian and Lindbladian governing the equations of motion in the single $\vec{k}_\parallel$-subspace to give (see Appendix~\ref{sec:appendix:GreenFormalism} and Ref.~\cite{lentrodt_ab_2020} for details)
\begin{align}
    \hat{H}=\frac{\omega_{\mathrm{nuc}}}{2}\opsigma^z_{\vec{k}_\parallel}+\Delta_\mathrm{CLS}{\opsigma}^+_{\vec{k}_\parallel}{\opsigma}^-_{\vec{k}_\parallel}
    -\left(\Omega{\opsigma}^+_{\vec{k}_\parallel}+\mathrm{h.c.}\right) \label{eq:single1}
\end{align} 
and 
\begin{align}
    \mathcal{L}[\rho]&=\frac{\Gamma_\mathrm{SR}+\gamma_0}{2}\left[2{\opsigma}^-_{\vec{k}_\parallel}\rho{\opsigma}^+_{\vec{k}_\parallel}-\acomm{{\opsigma}^+_{\vec{k}_\parallel}{\opsigma}^-_{\vec{k}_\parallel}}{\rho}\right]\,, \label{eq:single2}
\end{align}
 respectively. Here, $\opsigma^{\pm}_{\vec{k}_\parallel}$ and $\opsigma^{z}_{\vec{k}_\parallel}$ are Pauli operators and $\Omega = \vec{d}^*\cdot\vec{E}_{\mathrm{in}}(\vec{k}_\parallel, z)N/{A}$ is the effective Rabi-frequency.

The equations~(\ref{eq:single1},~\ref{eq:single2}) indeed have the form of 
a quantum mechanical two-level system. However, the excited state is now a collective spin-wave excitation in the ensemble of nuclei embedded in the cavity, rather than one of the single-nucleus excitations considered in the many-body equations~(\ref{eq:many1},~\ref{eq:many2}). Furthermore, 
the properties of the effective two-level system are different compared to the bare nuclei. First, an additional detuning term $\Delta_\textrm{CLS}$ appears, which shifts the transition energy of the  two-level system, and which is known as the \textit{Collective Lamb Shift}~\cite{friedberg_frequency_1973, scully_collective_2009, ruostekoski_emergence_2016, keaveney_cooperative_2012, guerin_light_2017, peyrot_collective_2018, roof_observation_2016, bromley_collective_2016, rohlsberger_collective_2010}. Second, the radiative decay rate is enhanced by the additional superradiant~\cite{dicke_coherence_1954,gross_superradiance_1982, scully_super_2009,garraway_dicke_2011,guerin_light_2017} contribution $\Gamma_\textrm{SR}$ as compared to the single-particle decay. Third, the light-matter coupling depends on the effective in-plane nuclear number density $N/A$, which appears as a result of the two-dimensional in-plane Fourier transformation. The explicit expressions for the aforementioned constants are discussed in Sec.~\ref{sec:couplings} below. Importantly, they can be tuned via the cavity structure and the angle of incidence of the x-rays. For this reason, we refer to the effective description as a tunable artificial two-level system.

\begin{figure}[t]
\includegraphics[width=0.9\columnwidth]{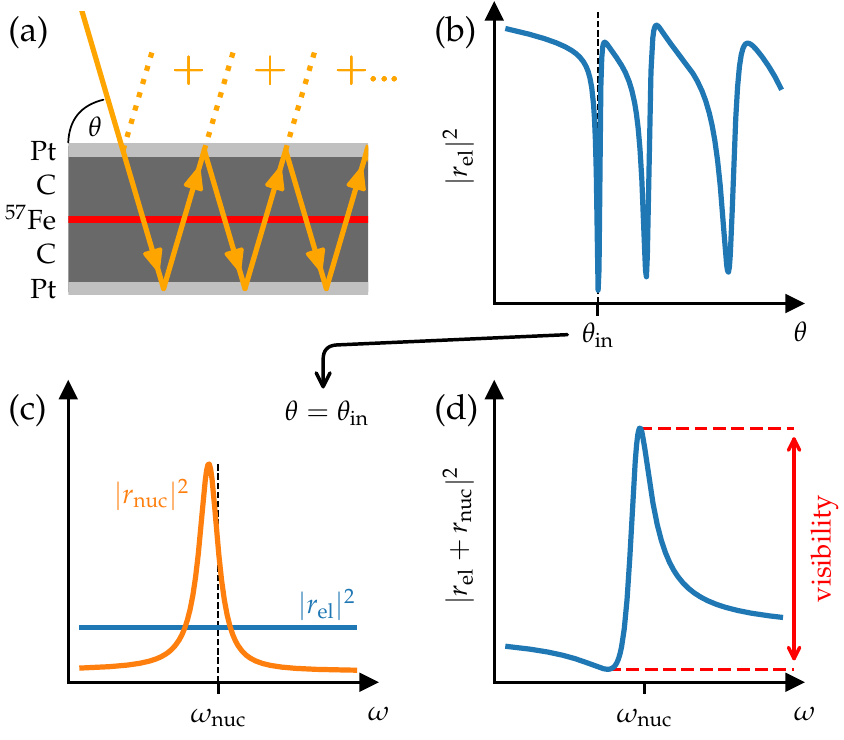}
\caption{(Color online) Cavity observables and visibility of the nuclear response in the reflection spectrum. (a) Schematic representation of the leading interfering contributions to the electronic cavity reflectance. For the actual calculation, reflections at the remaining layers, e.g. the \Fe layer, have to be taken into account as well. (b) Electronically reflected intensity (``rocking curve'') as calculated by the contributions in (a) using Parratt's formalism~\cite{parratt_surface_1954}. (c) Electronic $r_\mathrm{el}$ and nuclear $r_\mathrm{nuc}$ contributions to the reflectance as function of the frequency $\omega$ for a fixed angle of incidence $\theta_\mathrm{in}$. (d) Full reflection spectrum given by the sum of both (complex-valued) contributions. The peak-to-peak amplitude of the resulting Fano-resonance is taken as the visibility criterion.} 
\label{fig:visbility} 
\end{figure}

\subsection{\label{sec:couplings}Coupling constants}

It remains to discuss the coupling constants entering the equations of motion (\ref{eq:many1},~\ref{eq:many2}) and (\ref{eq:single1},~\ref{eq:single2}) in the many-body and the spin-wave basis, respectively. Within macroscopic QED, these constants can be expressed in terms of the Green's function characterizing the cavity environment~\cite{buhmann_dispersion_2012,asenjo-garcia_atom-light_2017}. For the present case of a layered dielectric medium, analytic expressions for the 
Green's function are derived in~\cite{tomas_green_1995}. 

In the many-body basis, the coupling and (cross-)decay constants evaluate to
\begin{align}
 &J_{nn'}={\mu_0\omega^2_{\mathrm{nuc}}}\vec{d}^*\cdot\Re\left[\vec{G}(\vec{r}_{n}, \vec{r}_{n'},\omega_\mathrm{nuc})\right]\cdot\vec{d}\,,\\
    &\Gamma_{nn'}=2{\mu_0\omega^2_{\mathrm{nuc}}}\vec{d}^*\cdot\Im\left[\vec{G}(\vec{r}_{n}, \vec{r}_{n'},\omega_\mathrm{nuc})\right]\cdot\vec{d}\,,
\end{align}
where $\mu_0$ is the vacuum permeability and $\vec{G}(\vec{r}_{n}, \vec{r}_{n'};\omega_\mathrm{nuc})$ the Green's function evaluated at the position of two nuclei $n,n'$ 
and the nuclear transition frequency $\omega_\mathrm{nuc}$.

In the effective description, the frequency shift $\Delta_\mathrm{CLS}$  and the enhancement of the spontaneous decay rate  $\Gamma_\textrm{SR}$ can be 
written as
\begin{align}
    \Delta_\mathrm{CLS}&=-\frac{N}{A}\mu_0\omega^2_{\mathrm{nuc}}\vec{d}^*\cdot\Re\left[\vec{G}(z, z, \vec{k}_\parallel,\omega_\mathrm{nuc})\right]\cdot\vec{d}\label{eq:DeltaEff}\,,\\
\Gamma_\mathrm{SR}&=2\frac{N}{A}\mu_0\omega^2_{\mathrm{nuc}}\vec{d}^*\cdot\Im\left[\vec{G}(z, z, \vec{k}_\parallel ,\omega_\mathrm{nuc})\right]\cdot\vec{d}
    \,, \label{eq:GammaEff}
\end{align}
where $\vec{G}(z, z, \vec{k}_\parallel;\omega_\mathrm{nuc})$ is the in-plane Fourier transformed electromagnetic Green's function related to the position-space Green's function via
\begin{equation}
\vec{G}(\vec{r}, \vec{r}')=\int\frac{\dd[2]{\vec{k}_\parallel}}{(2\pi)^2}\:\vec{G}(z, z', \vec{k}_\parallel)\:e^{i\vec{k_\parallel}\cdot(\vec{r}_\parallel- \vec{r}'_\parallel)}\,.
\end{equation}
Both remaining spatial arguments are evaluated at the nuclear layer depth $z$. 
Explicit expressions for the Green's function for the archetype cavity considered in this paper are provided in Appendix~\ref{sec:appendix:GFs}.

Finally, we discuss the driving field $\vec{E}_{\mathrm{in}}(\vec{k}_\parallel, z)$ appearing in Eq.~(\ref{eq:single1}). It relates  the in plane Fourier transform to its real-space representation, 
\begin{equation}
\vec{E}_{\mathrm{in}}(\vec r)=\int\frac{\dd[2]{\vec{k}_\parallel}}{(2\pi)^2}\:\vec{E}_\mathrm{in}(z, \vec{k}_\parallel)\:e^{i\vec{k_\parallel}\cdot\vec{r}_\parallel}\,.
\end{equation}
Inside the cavity, the  externally applied field is modified due to absorption and reflection by the cavity materials. Quantitatively, its frequency space solutions can be obtained, e.g., using Parratt's formalism~\cite{parratt_surface_1954}. For the system at hand, we give the explicit form of the field inside the cavity in Appendix~\ref{sec:appendix:fields}. Note that this field is to be calculated without considering the nuclear resonance, but including the electronic index of refraction of the nuclear layer.

\subsection{Observables\label{sec:theory:observables}}

\subsubsection{Reflection spectrum}
The key observable for nuclei embedded in thin-film cavities, dominating the experimental work up to now, is the linear spectrum of the reflected light 
measured for a fixed incidence angle of the probing x-rays, see Fig.~\ref{fig:visbility}. Following Ref.~\cite{lentrodt_ab_2020}, we summarize the relevant aspects of this observable in the present context. 

Noting that the Fresnel coefficients  for $s$- and $p$-polarization become equivalent at grazing incidence~\cite{als-nielsen_elements_2011}, and that the electronic scattering and the scattering on the single unsplit nuclear resonance leave the polarization of the incident x-rays unchanged, 
we subsequently focus on the treatment of s-polarized light and omit the vectorial nature of the electric field and Green's function. 

The reflection spectrum comprises contributions by the purely electronic reflection at the different layer boundaries and by the artificial nuclear two-level system. We illustrate these in Fig.~\ref{fig:visbility}. 
We calculate the purely electronic reflection $r_\mathrm{el}$ using Parrat's formalism~\cite{parratt_surface_1954}, which sums all the different scattering contributions arising from the material boundaries. Although in principle the electronically reflected light  has a frequency dependence, it can be neglected on scales of the linewidth of the nuclei. 
The artificial two-level system gives rise to a Lorentzian spectrum in linear response. However, the nuclear response acquires an additional complex-valued weight upon propagation to the cavity surface, and interferes with the electronically reflected background, which is complex-valued. Details on the calculation are found in Appendix~\ref{sec:appendix:GreenFormalism}.

In combination, the overall reflected intensity normalized to the incoming intensity is given by  
\begin{align}
\left|r(\vec{k}_\parallel, \omega)\right|^2 = \left||r_\mathrm{el}|+\frac{|A|e^{i\varphi}}{\Delta+i\Gamma}\right|^2\,.
\label{eq:theory:Fano}
\end{align}
Depending on the relative phase $\varphi$ of both contributions, different nuclear Fano lineshapes arise in the spectra~\cite{ott_lorentz_2013,limonov_fano_2017, vassholz_observation_2021, heeg_interferometric_2015}. The response is centered at the transition frequency of the collective two-level system and superradiantly broadened, such that
\begin{align}
\Delta &= \omega-(\omega_\mathrm{nuc}+\CLS)\,,  \\
\Gamma&=(\gamma_0+\SR)/2 \,.
\end{align}
The relative weight of the nuclear contribution $A$, resulting from the coupling of the nuclei to the driving field as well as the propagation of the nuclear response to the cavity surface,
can be expressed as
\begin{align}
&A = -\mu_0\omega_\mathrm{nuc}^2|{d}|^2{G}(0, z, \vec{k}_\parallel, \omega_\mathrm{nuc}){E}_\mathrm{in}(z, \vec{k}_\parallel, \omega_\mathrm{nuc}) \,,\label{eq:theory:paramsReflA}
\end{align}
again using the Green's function. Finally, the relative phase between these two contributions determining the line shape of the nuclear resonance is~\cite{ott_lorentz_2013,limonov_fano_2017,heeg_interferometric_2015} 
\begin{align}
&\varphi=\arg(A)-\arg(r_\mathrm{el})\,.
\label{eq:theory:paramsReflPhi}
\end{align}
Explicit expressions for the Green's function, the electric field configuration and the electronic cavity reflection are provided in Appendix~\ref{sec:appendix:formulae}.

\subsubsection{Visibility of the nuclear response in the reflection spectrum}\label{sec:theory:visibility}
From Eqs.~\eqref{eq:theory:Fano} and \eqref{eq:theory:paramsReflA} it is clear that the nuclear signatures in the reflection spectrum can be strongly suppressed, e.g., by the Green's function contribution. One obvious reason for strong attenuation are thick or highly absorptive cavity layers. 
This poses the problem that the inverse design may lead to optimized solutions that in practice cannot be observed via the reflection spectrum. In such situations, the formally best cavity structures may not be the most relevant ones for experimental purposes. This challenge can be tackled by including conditions on the practical relevance, such as the observability,  into the design rules. 

Considering the lineshape of a Fano-resonance, a suitable criterion for the visibility of the nuclear signatures is the peak-to-peak amplitude of the resonance in the normalized reflection spectrum, see. Fig.~\ref{fig:visbility}(d). In the optimization, one may then set a minimum visibility as a boundary condition, or optimize the visibility for an otherwise specified design goal.

An efficient calculation of the visibility criterion is possible via analytical expressions for the positions of the two extrema of Eq.~\eqref{eq:theory:Fano} in terms of $\Delta$, 
\begin{align}
\Delta_{\pm}=&-\frac{1}{2|r_\mathrm{el}|}\Big[ |A|\sec(\varphi)+2|r_\mathrm{el}|\Gamma\tan(\varphi)\notag\\ \pm &\sec(\varphi)\sqrt{|A|^2+4|r_\mathrm{el}|^2\Gamma^2+4|A||r_
\mathrm{el}|\Gamma\sin(\varphi)}
\Big]\,.
\end{align}
In part of the following calculations, we will use this visibility criterion as  an additional design constraint.

\begin{figure*}[t]
\includegraphics[width=\textwidth]{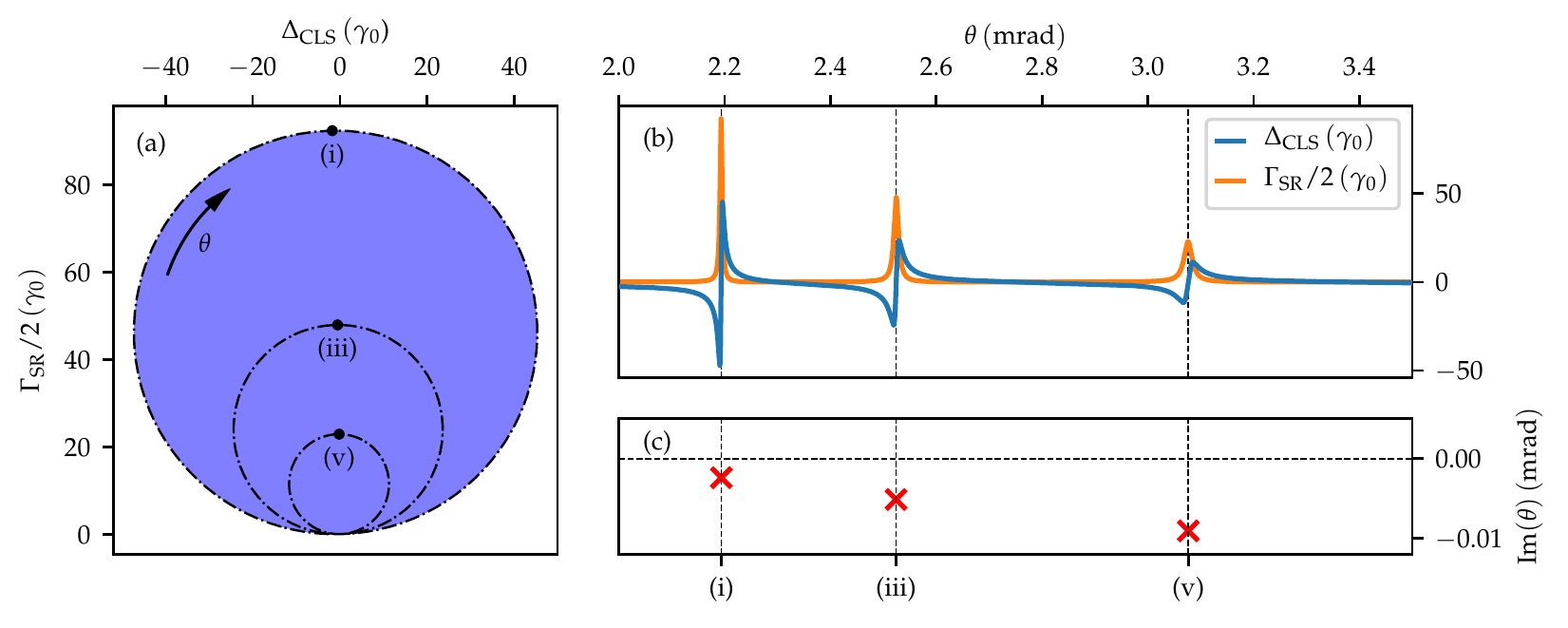}
\caption{(Color online) Accessible parameter space for the resonance frequency shift \CLS and the line width broadening \SR of the artificial x-ray 
two-level system. The figure shows the archetype cavity with Pt/C/\Fe /C/Pt/Si structure, as illustrated in Fig.~\ref{fig:schematicSetup}. (a) The 
blue shaded area indicates the accessible combinations of $\CLS$ and $\SR$. Starting from the topmost point with highest SR, the outermost, black dashed circle is traversed upon tuning the angle of incidence around the first cavity resonance. The circles with smaller radii are accessed by tuning the incidence angles around the higher cavity modes. (b) Explicit parametrization of the trajectory in (a) in terms of $\CLS$ and $\SR$ via the angle of incidence. (c) shows the poles of the Green's function against the angle of incidence, which can be associated with the individual cavity modes. The dots on the curve in (a), labeled with lower-case roman numerals,  relate the structures in (a) to the poles in (d). The cavity structure corresponding to the circles in (a) and the results in (b,~c) is Pt($80.4\,\mathrm{nm}$)/C($46.0\,\mathrm{nm}$)/\Fe{}($0.57\,\mathrm{nm}$)/C($46.1\,\mathrm{nm}$)/Pt($17.8\,\mathrm{nm}$)/Si.}
\label{fig:CLS_SR_TrajectoryPoles}
\end{figure*}

\section{\label{sec:CLS-SR-VIS}Inverse Design of the artificial two-level system}

For the archetype cavity system we explore the combinations of CLS and SR that are in principle accessible. Once the accessible parameter space is known, at least one cavity structure can be associated to each case, thereby allowing for the inverse design of the artificial two-level system.

For the discussion we employ the Pt/C/\Fe /C/Pt/Si cavity of Fig.~\ref{fig:schematicSetup}, similar to cavities commonly used in experiments. The resonant layer, i.e. the layer containing the nuclear resonances, is chosen to be about two atomic layers of \Fe which corresponds to a layer thickness of $0.574\,\mathrm{nm}$, as has been utilized in~\cite{rohlsberger_collective_2010}.  At such low layer thicknesses, long-range magnetic order  and magnetic hyperfine splittings are suppressed, such that the iron nuclei can be approximated as unsplit two-level systems. The other layer thicknesses as well as the angle of incidence remain as tuning parameters.

We explore the accessible quantum optical parameters by scalar minimization routines within the \texttt{scipy.optimize}~\cite{scipy_10_contributors_scipy_2020} package in python. To explore parameter spaces beyond one dimension, we extremize suitable linear and non-linear scalar combinations of the observables. Details on the methods applied can be found in Appendix~\ref{sec:appendix:numericalMethods}.

\subsection{\label{sec:CLS_AND_SR} Frequency shift and decay enhancement as the design goals}

Results for the CLSs and SRs in the archetype cavity of Fig.~\ref{fig:schematicSetup} are shown in  Fig.~\ref{fig:CLS_SR_TrajectoryPoles}(a). The blue-shaded area indicates the combinations of $\Delta_\mathrm{CLS}$ and $\Gamma_\mathrm{SR}$ that can be realized. Interestingly, we find this set to be circular. Enhanced spontaneous emission is found everywhere, except at one point, whereas the CLS can take positive, zero, as well as negative values. The whole circle is slightly shifted to negative CLS. Being able to give explicit cavity geometries for each individual point within the set, we achieve the basis for the inverse design of artificial two-level schemes.

In order to explain the highly symmetric set of CLS and SR, we can consider a cavity at the boundary of the blue circle in  Fig.~\ref{fig:CLS_SR_TrajectoryPoles}(a), e.g., the cavity featuring highest SR. Fixing this cavity structure and tuning the angle of incidence around the first cavity resonance, the outermost black dashed trajectory indicated in Fig.~\ref{fig:CLS_SR_TrajectoryPoles}(a), is traversed clockwise with increasing incidence angle. Increasing the angle further towards the next-higher cavity resonance, the second-largest circle indicated in the figure is traversed, and so on. 
The explicit parametrization of this trajectory with the angle of incidence is shown in Fig.~\ref{fig:CLS_SR_TrajectoryPoles}(b). For each resonance structure, indicated by a peak in the SR and a zero in the CLS, a circle of different radius is traversed. Since  the circle with largest diameter constitutes the boundary of the accessible CLSs and SRs, we find that the highest possible CLSs and SRs can be achieved within a single cavity geometry. This is consistent with previous predictions~\cite{longo_tailoring_2016}.

\begin{figure*}[t]
\includegraphics[width=0.95\textwidth]{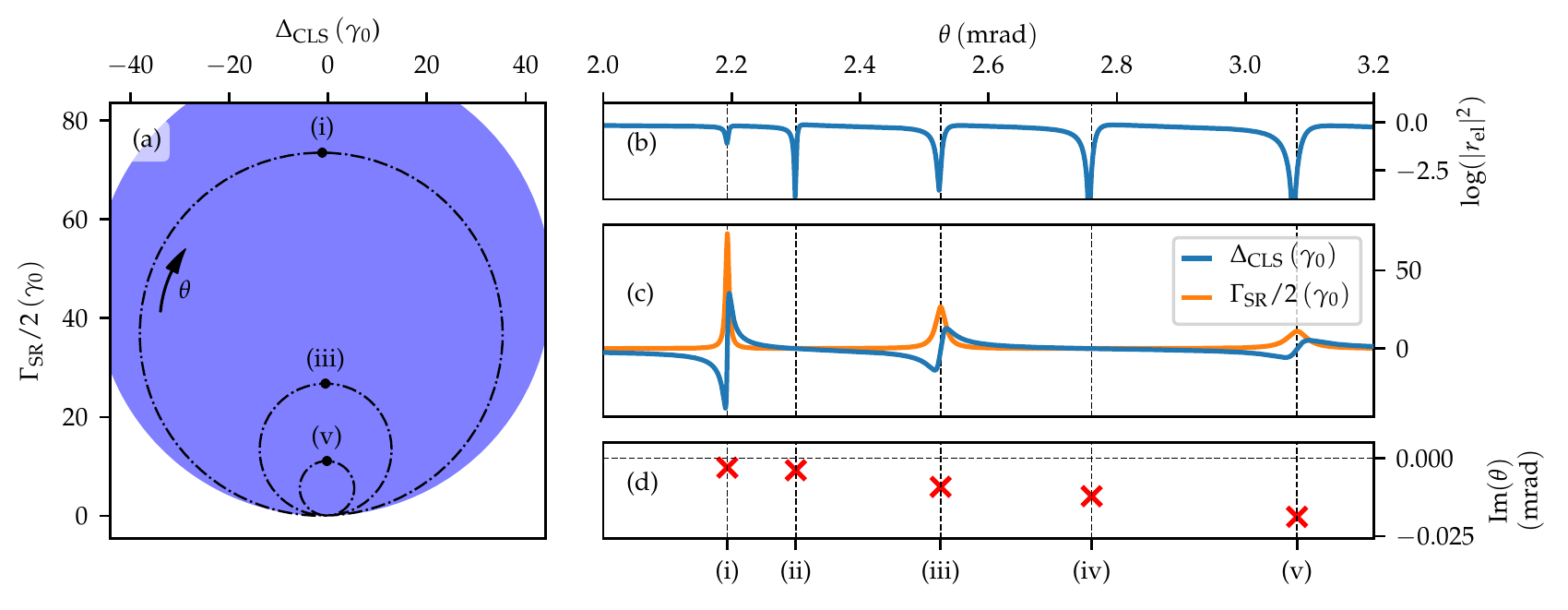} 
\caption{(Color online) Characterization of a cavity setup with the additional constraint of a visibility of the nuclear Fano line signature in the reflection spectrum given by 50\% of the incident x-ray intensity, again realized in a Pt/C/\Fe /C/Pt/Si cavity. Panels (a), (c) and (d) are analogous to the corresponding panels in Fig.~\ref{fig:CLS_SR_TrajectoryPoles}. Panel (b) further shows the reflected intensity (``rocking curve'') as function of the angle of incidence $\theta$. The cavity structure corresponding to the circles in (a) and the results in (b-d) is Pt($2.7\,\mathrm{nm}$)/C($45.7\,\mathrm{nm}$)/\Fe{}($0.57\,\mathrm{nm}$)/C($46.1\,\mathrm{nm}$)/Pt($307.3\,\mathrm{nm}$)/Si.}

\label{fig:Trajectory_poles_mediumVis}
\end{figure*}

Fig.~\ref{fig:CLS_SR_TrajectoryPoles}(c) shows the pole structure of the Green's function as a function of the angle of incidence. Noting that the resonant layer is placed precisely in the center of the guiding layer for the cavity considered in Fig.~\ref{fig:CLS_SR_TrajectoryPoles}, the thick, absorptive cladding layers on both sides of the guiding layer effectively ensure a mirror-symmetry around the resonant layer. Therefore, the odd parity modes in the guiding layer feature nodes at the nuclear layer and only even parity modes can couple to the nuclei. For this reason we leave out every second roman numeral for the labeling of the poles in Fig.~\ref{fig:CLS_SR_TrajectoryPoles}.

Notably, we find that each circle in Fig.~\ref{fig:CLS_SR_TrajectoryPoles}(a) is associated to a respective pole in panel (c). The circles are traversed in a continuous way upon passing by the corresponding poles.
To understand this behaviour, we can express the Green's function by a  Mittag-Leffler pole expansion~\cite{arfken_mathematical_2013, lalanne_light_2018,lentrodt_classifying_2021}  in the angle of incidence $\theta$ at constant frequency, i.e. we write it as
\begin{equation}
G(\theta) = G(\theta = 0) + \sum\limits_{\theta_0}\mathrm{Res}(G, \theta_0)\left(\frac{1}{\theta_0}+\frac{1}{\theta-\theta_0}\right)\,,
\label{eq:MLGreen}
\end{equation}
where $\theta_0$ are the poles of the Green's function and $\mathrm{Res}(G, \theta_0)$ the respective residua. Each pole can be associated to a cavity mode coupling to the resonant layer~\cite{lentrodt_classifying_2021}. The imaginary part of the pole then sets the width of the respective mode. For the cavity at hand, Fig.~\ref{fig:CLS_SR_TrajectoryPoles}(c) indicates that the imaginary parts of the poles are very small as compared to their real part separation. Being close to one pole thus allows to accurately describe the Green's function by a single-mode approximation,
\begin{equation}
G_\mathrm{SM}(\theta) = \mathcal{C}+\frac{\mathrm{Res}(G, \theta_0)}{\theta-\theta_0}\,,
\label{eq:ML_SM}
\end{equation}
where $\mathcal{C}$ accounts for the relevant terms constant in $\theta$. 
Upon tuning $\theta$, the expression maps to a circle in the complex plane spanning CLS and SR. This is in accordance with the single-mode approximation being a M\"obius transformation. The residues and imaginary parts may vary among the different poles, which explains the distinct radii for the first three modes, visible in the trajectory in Fig.~\ref{fig:CLS_SR_TrajectoryPoles}(a).

The small imaginary part of the poles is understood when realizing  that the cavity design chosen for this discussion (i.e. the one at the boundary of the circle) features very thick cladding layers. This is not surprising since the latter raise the intra-cavity reflectivity and hence form the basis for stronger inter-nuclear couplings, thus accounting for larger 
collective effects. Likewise, larger intra-cavity reflectivity allows for 
more narrow modes which explains the poles' behaviour.

\section{\label{sec:inv-vis}Inverse design with the visibility as an additional design goal}

\begin{figure}[t]
\includegraphics[width=0.9\columnwidth]{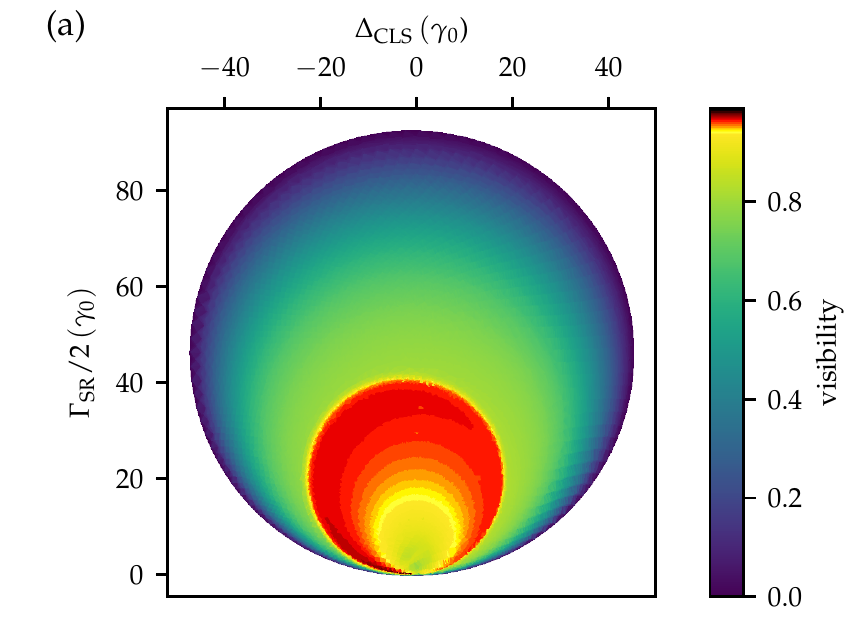}
\includegraphics[width=0.9\columnwidth]{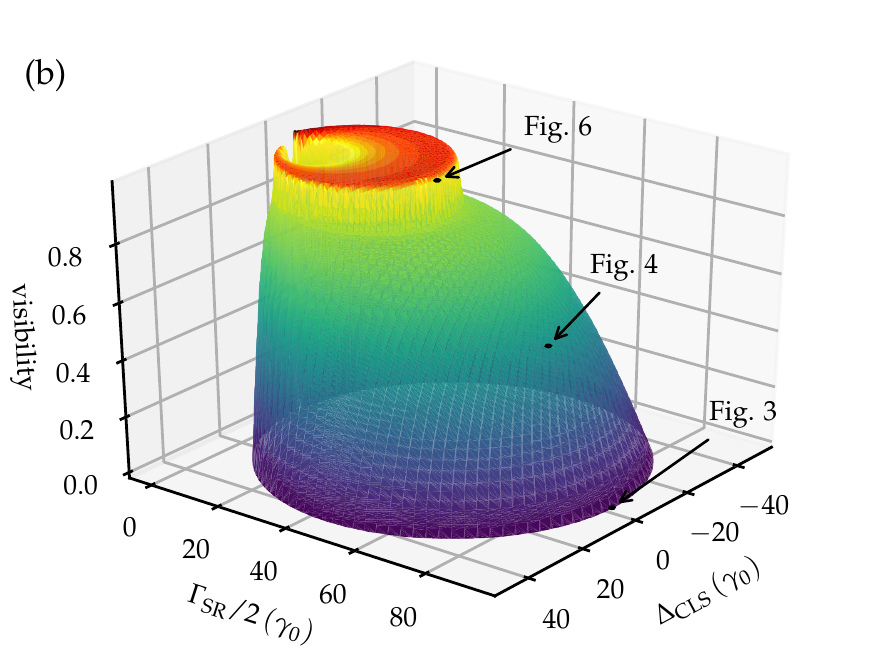}
\caption{(Color online) Accessible parameter space for the artificial x-ray two-level system with the three design goal parameters $\SR$, $\CLS$ and visibility. Results are shown for Pt/C/\Fe /C/Pt/Si cavity structures. (a) For each accessible $\CLS$ and $\SR$, the visibility is maximized, and shown via the color-coding. (b) Three dimensional representation of the parameter space in (a), which shows a plateau of high visibility as a distinct parameter region. Note that the accessible parameter combinations are not restricted to the surface only, but also encompass combinations inside the shown structure. The cavities used for Figs.~\ref{fig:CLS_SR_TrajectoryPoles}, \ref{fig:Trajectory_poles_mediumVis} and \ref{fig:Trajectory_poles_plateau} are marked in the figure.}
\label{fig:CLS_SR_VIS}
\end{figure}

In the preceding Section \ref{sec:CLS_AND_SR}, we found that the highest possible CLS and SR at the circle's boundary in Fig.~\ref{fig:CLS_SR_TrajectoryPoles}(a) are realized in a cavity with thick cladding layers. However, while this increases the cavity-mediated exciton self-coupling and thereby enhances the collective effects, it at the same time suppresses the coupling of light into and out of the cavity mode.  Practically, this means that the associated artificial two-level schemes cannot be observed via the reflectance with high visibility in experiments. In order to quantify this practical restriction, we add the visibility, defined in section \ref{sec:theory:visibility}, 
as a third observable characterizing the cavity performance. 

As a first step towards practically relevant settings, we consider cavities with a visibility of about 50\%, i.e. the reflection spectrum is modulated by half the intensity impinging on the cavity. Here, we once again search for the cavity realizing highest SR possible. Again fixing the different layer thicknesses and varying the angle of incidence, Fig.~\ref{fig:Trajectory_poles_mediumVis} shows the characteristics of this cavity, in a representation analogous to Fig.~\ref{fig:CLS_SR_TrajectoryPoles}. Owing to the considerably reduced top cladding layer thickness ($2.7\,\mathrm{nm}$), the SR accessible in this cavity is clearly lower, however, still significant. In addition to the previous panels, we also consider the intensity as reflected on the cavity in Fig.~\ref{fig:Trajectory_poles_mediumVis}(b) which has not been a meaningful observable in Fig.~\ref{fig:CLS_SR_TrajectoryPoles} because of the high top-layer thickness leading to a vanishing visibility of the nuclear signature. Comparing Fig.~\ref{fig:Trajectory_poles_mediumVis}(b) and (c), we see that only every second mode couples to the resonant layer. Although the poles in the Green's function are in principle present for modes (ii) and (iv), their numerical effect in Fig.~\ref{fig:Trajectory_poles_mediumVis}(c) is negligible due to their odd parity resulting in nodes at the resonant layer. While for higher order modes the minima of the reflected intensity Fig.~\ref{fig:Trajectory_poles_mediumVis}(b) do not coincide with the poles' real parts due to overlapping modes, these multi-mode effects~\cite{lentrodt_classifying_2021} are suppressed in the Green's function since relevant modes have roughly twice the distance. Thus, the Green's function can still be treated by the single mode expression Eq.~\eqref{eq:ML_SM} and we find circles in Fig.~\ref{fig:CLS_SR_VIS}(a).

To arrive at a comprehensive description of the relation between visibility and quantum optical parameters of the two-level scheme, we subsequently optimize for the highest possible visibility given some combination of CLS and SR in the accessible parameter space. The result of this optimization is shown in Fig.~\ref{fig:CLS_SR_VIS}(a). Note that the ring-like structures visible in this plot 
do not correspond to the seemingly similar structures in Fig.~\ref{fig:CLS_SR_TrajectoryPoles}(a) and Fig.~\ref{fig:Trajectory_poles_mediumVis}(a). While the former are obtained with different cavity structures optimizing the visibility, the latter are drawn for a single cavity structure. Likewise, Fig.~\ref{fig:CLS_SR_VIS}(b) shows a three-dimensional representation of the overall possible combinations of CLS, 
SR and visibility. Clearly, as discussed above, the maximum SR and CLS result in near-zero signature in the reflectance. As we approach the interior of the circular set, the accessible visibilities become larger, and eventually and indeed quite abruptly, a saturation to values very close to one (color-coded red in Fig.~\ref{fig:CLS_SR_VIS}) is observed, while still allowing for comparatively high SR and CLS.

We can explain this peculiar abrupt saturation by a qualitative change in the optimum geometry of the cavity. Counter-intuitively, the optimum cavities giving rise to this plateau of highest visibilities do not have any upper cladding layer. This is unexpected, as such structures are more similar to single-mirror settings, rather than cavities. 

\begin{figure*}[t]
\includegraphics[width=0.95\textwidth]{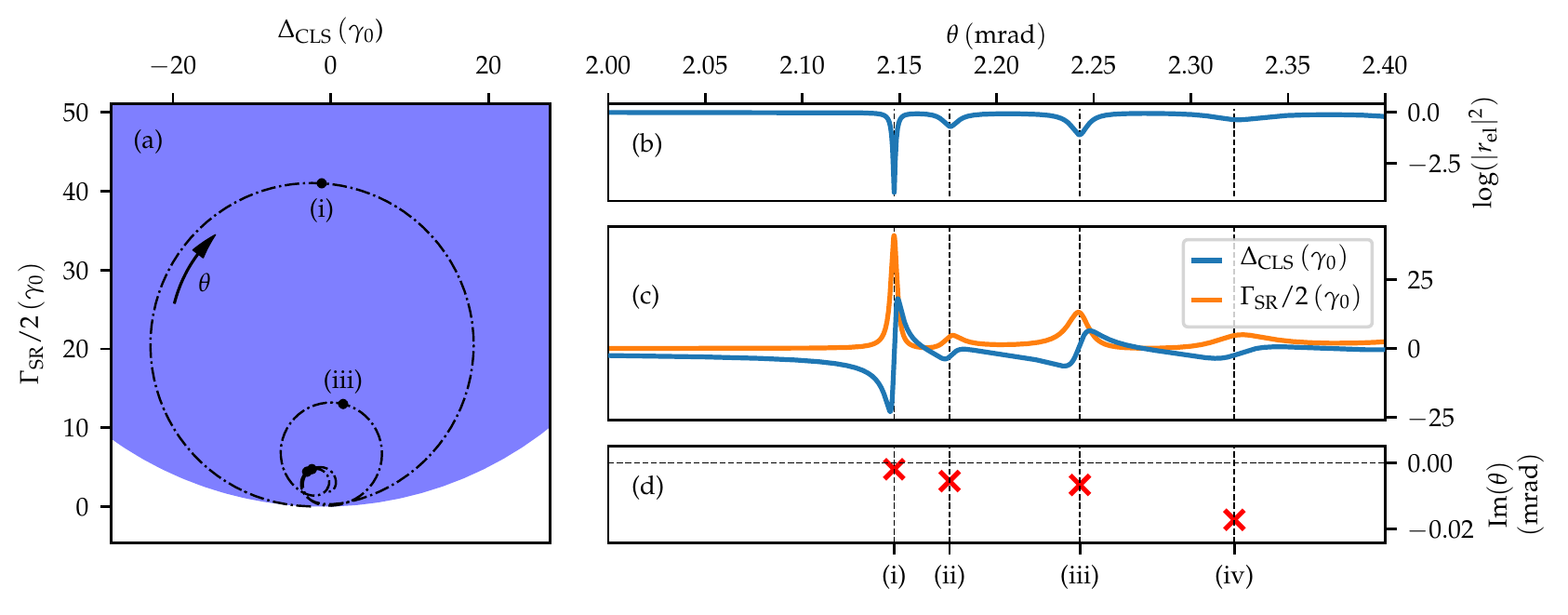}
\caption{(Color online) Characterization of a C($80.1\,\mathrm{nm}$)/\Fe{}($0.57\,\mathrm{nm}$)/C($102.6\,\mathrm{nm}$)/Pt($17.6\,\mathrm{nm}$)/Si cavity-setup, corresponding to a two-level system with high visibility of about 94\%. This cavity is marked in the global design parameter space shown in Fig.~\ref{fig:CLS_SR_VIS}. The panels are analogous to those in  Fig.~\ref{fig:Trajectory_poles_mediumVis}.}
\label{fig:Trajectory_poles_plateau}
\end{figure*}

We illustrate the performance of such a cavity without top cladding in Fig.~\ref{fig:Trajectory_poles_plateau} (cf. indication in Fig.~\ref{fig:CLS_SR_VIS}). For the specific setting, the SR takes values up to about $40\,\gamma_0$ while maintaining a large visibility. In contrast to the previous examples Fig.~(\ref{fig:CLS_SR_TrajectoryPoles},~\ref{fig:Trajectory_poles_mediumVis}), the cavity is not symmetric anymore and the coupling of different modes to the nuclei in the resonant layer in Fig.~\ref{fig:Trajectory_poles_plateau}(c) does not follow a simple pattern as before. Prominently, the incoupling into the first mode and the coupling of the first mode to the resonant layer are most pronounced. Since neighbouring modes are not well-separated on the scale of their widths, the single-mode approximation of Eq.~\eqref{eq:ML_SM} is not applicable. In particular for higher order modes, the trajectories in Fig.~\ref{fig:Trajectory_poles_plateau} thus loose their circular appearance and become spiral-like as also nearby poles contribute.  

For future experiments, the plateau in Fig.~\ref{fig:CLS_SR_VIS} points to a new, possibly preferential approach since cavities without 
upper cladding provide very clear spectral signatures while still showing 
significant collective effects. 

\begin{figure*}[t]
\includegraphics[width=0.95\textwidth]{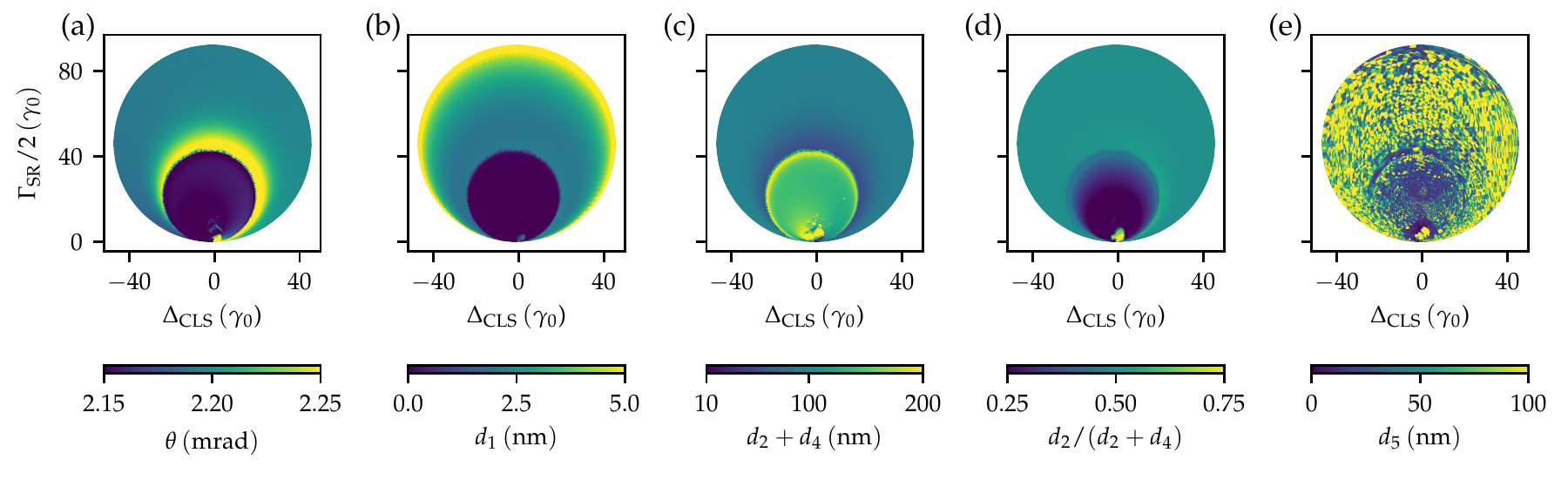} 
\caption{(Color online) Inverse design of the artificial two-level system. Results are shown for Pt/C/\Fe /C/Pt/Si cavity systems. The panels show cavity parameters that allow for the design of the two-level systems of optimal visibility, see Fig.~\ref{fig:CLS_SR_VIS}. These include (a) the angle of incidence, (b) the top cladding layer thickness, (c) the guiding layer thickness, (d) the relative position of the resonant layer in the guiding layer and (e) the bottom cladding layer thickness. See Fig.~\ref{fig:schematicSetup} for the illustration of the cavity parameters.}
\label{fig:InverseDesign}
\end{figure*}

With the foregoing discussion we are not only able to give precise cavity 
structures for the realization of a desired quantum optical two-level scheme, but can also quantify to what extent it will be visible in the reflectance. To showcase the successful inverse design, Fig.~\ref{fig:InverseDesign} presents the cavity parameters that constitute the surface of optimal visibility at fixed CLS and SR in Fig.~\ref{fig:CLS_SR_VIS}. The cavities are characterized by five parameters which are shown in Fig.~\ref{fig:InverseDesign}(a)-(e). For each desired parameter set of the artificial two-level system, the corresponding cavity design can directly be read off. 
From Fig.~\ref{fig:InverseDesign}(e) we clearly see that the bottom cladding layer thickness is not described by a function continuous in the CLS and SR. The reason for this is that the design constraints used to obtain the results do not uniquely fix the bottom cladding layer thickness. This can be understood by the fact that from a certain thickness on, the system  for all practical purposes is indifferent to a further increase in this thickness as the transmission through the bottom cladding is suppressed exponentially. Figure~\ref{fig:InverseDesign} also suggests that for most applications the usage of a thick bottom cladding is preferential. For the remaining parameters we find mostly continuous dependencies on the CLS and SR, which is somewhat unexpected as only three external constraints (CLS, SR and visibility) were imposed and our numerical procedure was not biased towards this continuous dependency. Overall, this completes the inverse design of artificial two-level schemes under the constraint of the nuclear response visibility in the reflection spectrum.

\section{\label{sec:CLS-SR-EF}Inverse design with the intra-cavity field enhancement as an additional design goal}
In the previous Sections, we found that the highest SR and CLS are realized in cavity structures with thick cladding layers, which do not allow for efficient in- and out-coupling of light, and therefore lack good visibility in the reflectance. In this section, we further explore this aspect by contrasting the CLS and SR with the intra-cavity enhancement of an external driving field at the resonant layer. These two design criteria are expected to lead to different optimal cavity designs, since the CLS and SR are maximized in the absence of coupling in- and out of the cavity mode, while the optimization of the intra-cavity enhancement of the externally applied field relies on a compromise between coupling of external fields to the cavity mode and enhancement of the light inside the cavity by multiple reflections.

First, we analyse the possible combinations of these quantities for the archetype cavity of Fig.~\ref{fig:schematicSetup} and comment on the peculiarities of this very setting. Subsequently, we proceed to compare the results to the behaviour expected for an optical  Fabry-P\'erot cavity and outline the conceptional differences.

\begin{figure}[t]
\includegraphics[width=0.9\columnwidth]{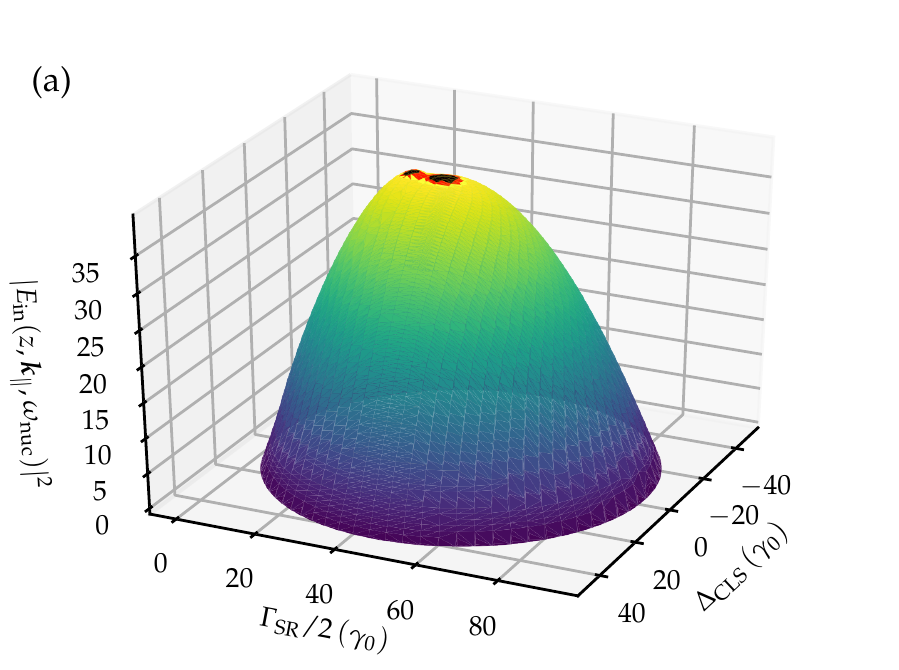} 
\includegraphics[width=0.9\columnwidth]{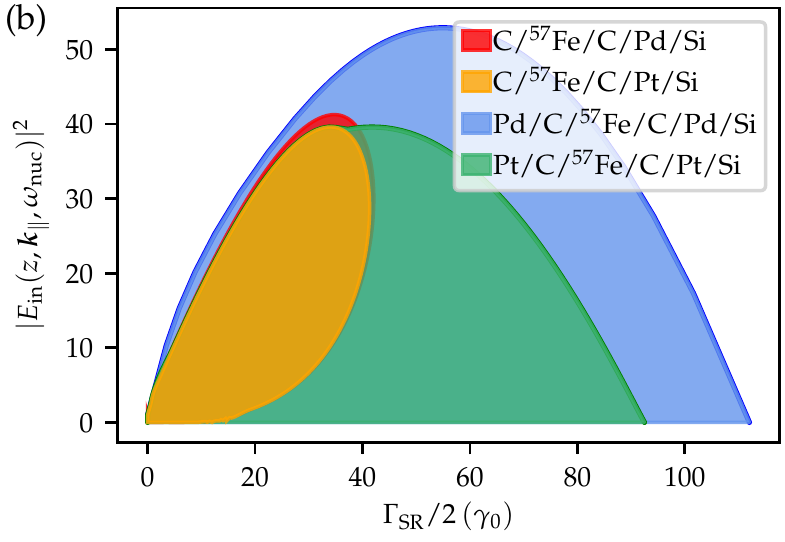}
\caption{(Color online) Accessible parameter space for the artificial x-ray two-level system with the cavity-induced field enhancement at the position of the nuclei as an additional design goal. (a) For each accessible pair of CLS and SR, the maximum possible field enhancement is obtained, and shown as a 3d plot with color coding. Combinations on and beneath the surface are accessible. The colouring highlights the double tip structure of the parameter space close to the highest possible field enhancements. This panel shows results for Pt/C/\Fe /C/Pt/Si cavity structures. (b) Section through panel (a) along the SR axis. The graph further compares the results for cavities with and without top cladding, as well as Pd and Pt as cladding materials, to explain the origin of the double-tip structure visible in (a).}
\label{fig:CLS_SR_EF_PtCPt}
\end{figure}

\subsection{Results for x-ray thin-film cavities}
The accessible combinations of CLS, SR and field enhancement at the nuclear layer are shown in Fig.~\ref{fig:CLS_SR_EF_PtCPt}(a). As expected, we find minimal field enhancement for the case of maximum CLS and SR. In going towards higher field enhancement, the maximum possible CLS and SR decrease. Interestingly, the set of possible parameter combinations features a double tip near the maximum field enhancements,  which is marked in black for clarity. To analyze its origin, we show a cut through panel (a) along the SR axis in Fig.~\ref{fig:CLS_SR_EF_PtCPt}(b) (green shaded area). This panel also shows corresponding results for a cavity without the topmost cladding layer (orange shaded area). One can clearly see that one of the two peaks can be attributed to cavities without the topmost layer. The second peak with similar maximal field enhancement is due to an entirely different cavity structure with a cladding layer.  As in Sec.~\ref{sec:CLS-SR-VIS}, we again find that cavities without upper cladding layer may form an equally good, or even superior, approach to designing x-ray layer structures.

To contextualize these observations, we finally consider cavities in which the cladding material is changed from Pt to Pd. Results are shown with (blue) and without (red) top cladding in Fig.~\ref{fig:CLS_SR_EF_PtCPt}(b). For this material combination, no double-tip appears, and the non-cladded system is outperformed by the system with topmost Pd layer. This indicates that the Pt/C cavity system is peculiar, but also that the material choice is of great significance.

But before we elaborate on the topic of different materials below, there is yet another striking feature visible in Fig.~\ref{fig:CLS_SR_EF_PtCPt}(b). 
The decay enhancement for a quantum system is linked to the photonic density of states (DOS) at its position~\cite{novotny_principles_2006}. Considering Fig.~\ref{fig:CLS_SR_EF_PtCPt}, we clearly see that the maximal field enhancement is  achieved in a different cavity geometry than the  maximal SR. This observation is counter-intuitive, especially considering that for standard optical cavities, one might expect the coincidence of these limits.  In the next section~\ref{sec:fp}, we illustrate this aspect further by comparing the x-ray thin-film cavities to Fabry-P\'erot cavities.

\subsection{Comparison with optical Fabry-P\'erot cavities\label{sec:fp}}
\begin{figure}[t]
\includegraphics[width=0.9\columnwidth]{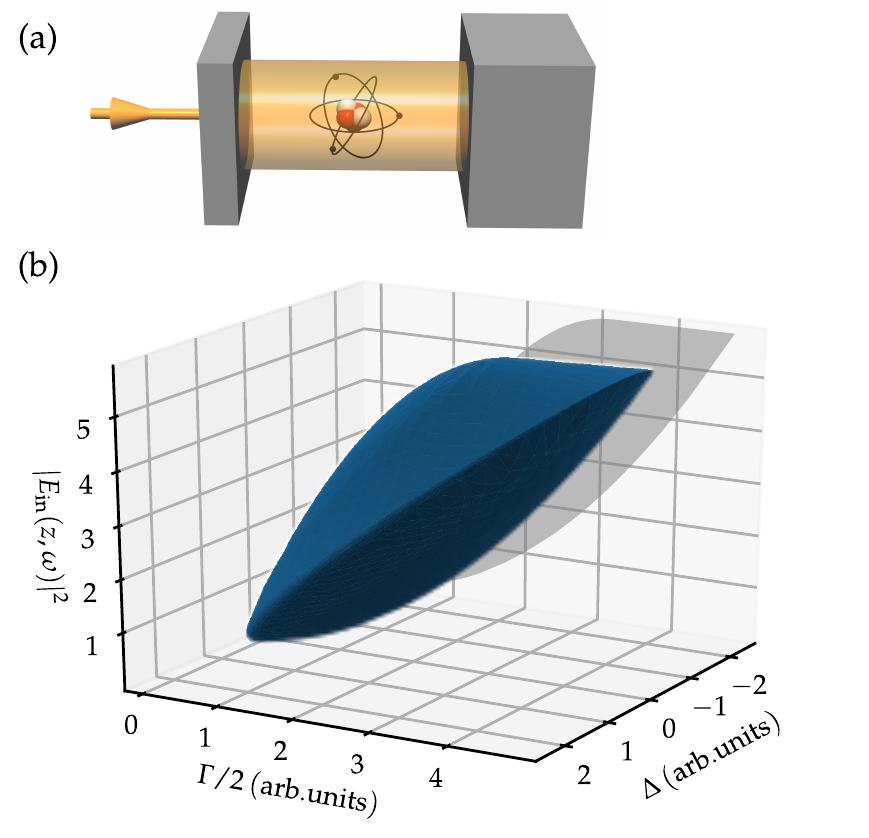}

\caption{(Color online) Accessible parameter space for a two-level atom in a Fabry-P\'erot cavity,  as a reference for comparison with the x-ray thin-film cavity case in Fig.~\ref{fig:CLS_SR_EF_PtCPt}. (a)~Schematic layout of the setup. The cavity is formed by diamond front and rear mirrors with rear thickness chosen large. A single two-level atom is 
coupled to the cavity field. The cavity is probed with light resonant to the atomic transition frequency. (b) Parameter space of accessible frequency shifts $\Delta$ and decay rate enhancements $\Gamma$ in the Fabry-P\'erot cavity. For each pair $\Delta, \Gamma$, the possible field enhancements at the atom's position $z$ are obtained, and shown as the third axis. The projection of the parameter space onto the $\Gamma-|{E}_{\mathrm{in}}(z, \omega)|^2$ plane is indicated. Unlike in the x-ray cavity case, the maxima of the line broadening $\Gamma$ and the field enhancement $|{E}_{\mathrm{in}}(z, \omega)|^2$ coincide.}
\label{fig:FabryPerot}
\end{figure}

To illustrate this qualitative difference, we analyse an one-dimensional lossy Fabry-P\'erot cavity as shown in Fig.~\ref{fig:FabryPerot}(a). The mirror material is chosen as diamond for its comparatively large refractive index of $n=2.4$, yet noting that variations of this refractive index only lead to quantitative, but not qualitative, changes. The cavity is illuminated at the resonance frequency of the TLS (chosen around $700\,\mathrm{nm}$) orthogonally to the mirror surface. For this setting, we seek the achievable combinations of frequency shift, decay enhancement and field enhancement at the TLS. It is readily described by the previously used formalism of Eqs.~(\ref{eq:Master}-\ref{eq:many2}) in the limit of a single atom. The coupling constants can be obtained from the Green's function in Appendix~\ref{sec:appendix:formulae}, which reduces to the one-dimensional real-space Green's function upon setting $\vec{k}_\parallel=0$.

To ensure comparability to the x-ray scenario, we not only vary the thicknesses of the layers and the position of the TLS therein, but also take the resonance frequency of the TLS as a parameter. Varying the frequency in this setting is tantamount to changing the angle in the x-ray case. 

Fig.~\ref{fig:FabryPerot}(b) shows the set of accessible combinations of field enhancement at the two-level system's position, level shift and decay enhancement. For clarity, we also show the projection of the set onto the SR-field enhancement-plane. Clearly, the maximal SR coincides with the largest possible field enhancement, as was suspected for this setting.

We can identify two major differences between the x-ray cavities and the Fabry-P\'erot setting: Firstly, x-ray materials tend to have high absorption compared to dielectrics available in the optical regime, and, secondly, the grazing incidence setup allows for total reflection and thus provides a different mechanism of trapping light. 
Regarding the second point, it is important to note that x-ray cavities as in Fig.~\ref{fig:schematicSetup} are operated below the angle of total reflection of the cladding layer, such that the in- and out-coupling will take place evanescently. Therefore, coupling of light into and out of the cavity may be suppressed by thick cladding layers even in the absence of absorption in the mirror material. In contrast, in this case, the photonic DOS may still be high, since it is limited essentially only by the absorption of the guiding layer. This illustrates why the field enhancement and the DOS are related in a different way than in the Fabry-P\'erot cavity case. Hence, we can attribute the counter-intuitive behavior of photonic DOS and field enhancement to the grazing incidence setup. Striving for higher finesse cavities, as is done in the visible domain, thus would go in line with neither driving nor observing the nuclear dynamics in our regime.  We note, however, that different in- and out-coupling schemes such as front-coupling~\cite{pfeiffer_two-dimensional_2002, jarre_two-dimensional_2005, chen_transient_2021} or intra-cavity generation of light~\cite{vassholz_observation_2021} could allow one to enhance the intracavity field enhancement  without reducing the SR and CLS.

Albeit the qualitative behavior of the parameter spaces is not affected by absorption, we nevertheless find it to exert striking influence on the 
quantitative performance of cavities. Therefore, the influence of guiding 
and cladding material properties is discussed in the following section.

\begin{figure*}[t]
\includegraphics[width=0.9\textwidth]{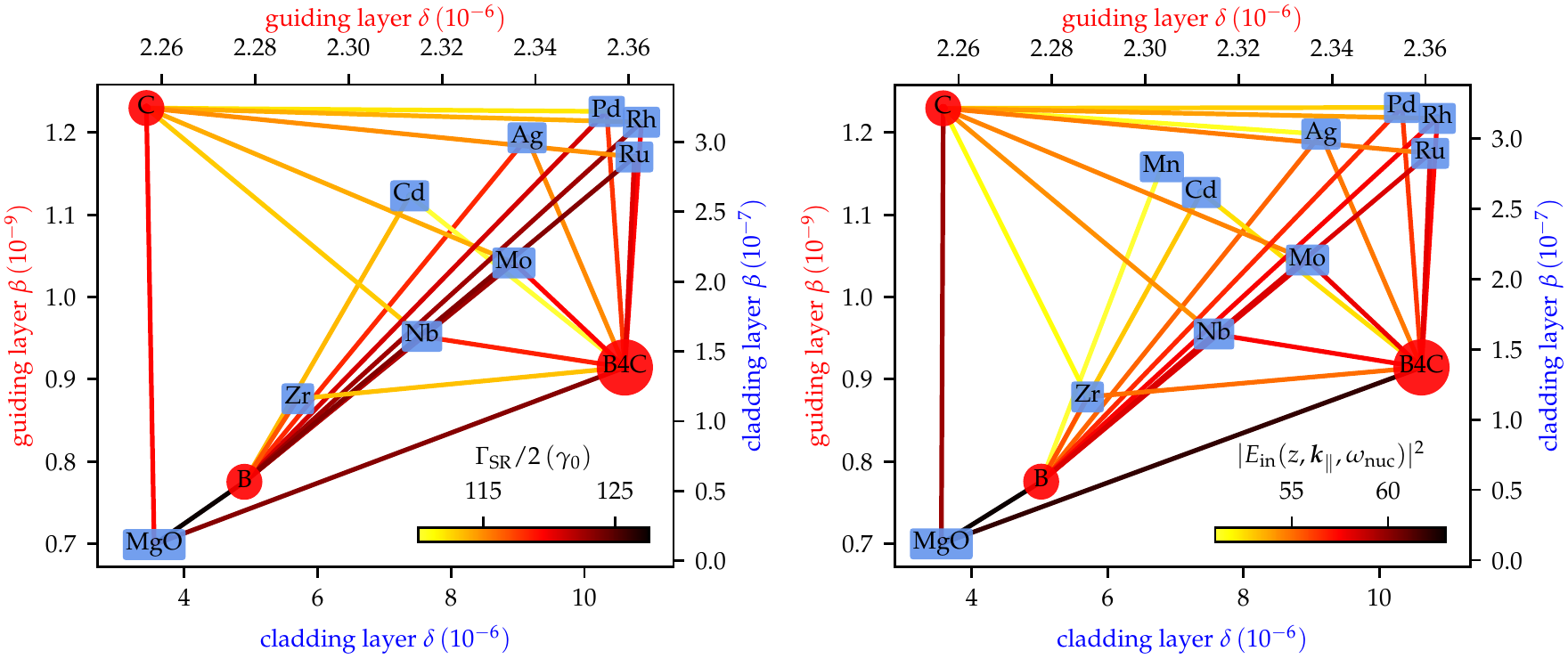}
\caption{(Color online) Survey of the tunability of the artificial x-ray two-level system as a function of the chosen layer materials. The left panel (a) shows the achievable SR, while the right panel (b) shows the field enhancement at the position of the nuclei. In each panel, the top [left] axes denote the real [imaginary] part of the refractive index $n=1-\delta + i\beta$ of the guiding layer material. Analogously, the bottom [right] axes characterize the refractive index of the cladding material. The 
different cladding [guiding] materials are shown as blue rectangular boxes [red circles] in the plot. Each line connecting a guiding-layer with a cladding-layer material defines a cavity structure. The optimum performance possible with this material combination 
is then color-coded in the line linking the two materials.}
\label{fig:Materials}
\end{figure*}

\section{\label{sec:Materials}Influence of the layer materials on the parameter space accessible by the inverse design}

X-ray materials are commonly found to have refractive indices close to and slightly below one, conventionally represented in the form~\cite{als-nielsen_elements_2011}
\begin{equation}
n=1-\delta+i\beta\,,
\end{equation} 
where $\delta, \beta$ are small, real numbers. 
For the design of x-ray thin-film cavities, the guiding principle commonly applied is taking \textit{high-$Z$}, i.e. high electron density, materials for the cladding and \textit{low-$Z$} materials for the guiding layer~\cite{rohlsberger_collective_2010,heeg_vacuum-assisted_2013,heeg_tunable_2015, haber_rabi_2017,yoshida_quantum_2021,adams_x-ray_2013,jaeschke_quantum_2020}. This guideline ensures high contrast in the real parts of the refractive indices and thus a comparatively large Fresnel reflectivity between adjacent layers. High electron densities, however, come along with high absorption which affects the cavity performance. The interplay between absorption and reflectivity of the cladding is not immediately clear, which is why we devote this section to the systematic analysis of the performance of different material combinations for the archetype cavity of Fig.~\ref{fig:schematicSetup}. 

To derive trends for the material choice, we sample the cladding and guiding materials from a range of elements as well as few chemical compounds known to be suitable for the manufacturing process. The \Fe-layer as well 
as the silicon substrate is left unchanged and the materials of the two cladding layers are taken to be the same. We note that related material samplings have  previously been reported in the context of bandpass filtering of broadband synchrotron radiation with grazing-incidence anti-reflection (GIAR) films~\cite{rohlsberger_grazing_1994}.

For each material combination we seek the highest possible field enhancement at the nuclear layer as well as highest possible SR. From the circular structures in Fig.~\ref{fig:CLS_SR_TrajectoryPoles}(a) and the pole expansion Eq.~(\ref{eq:ML_SM}) it can be expected that the cavities optimizing the SR also optimize the CLS. The best extremization 
outcomes for the SR and the field enhancement are shown in Fig.~\ref{fig:Materials}(a) and (b), respectively.

As expected, we find that the best cavities feature low absorption in the guiding layers. For the cladding materials, however, common materials such as Pt are neither among the best cavities for SR nor for the field enhancement. This is unexpected, since Pt is a high-$Z$-material with comparatively high $\delta$ which gives rise to a high Fresnel reflectivity at the cladding-guiding boundary. Overall, we find that Pt is not an exception, because the best cavites are not achieved via a high contrast in the refractive indices' real parts, but rather with the low absorption in the cladding as a selection criterion. Most strikingly, all metals are outperformed by the MgO compound as cladding material. 

These results indicate that the 
paradigm of \textit{high-$Z$}-cladding -- \textit{low-$Z$}-guiding material as a cavity design criterion has to be reconsidered. Instead, implementing \textit{low-$Z$}-cladding 
-- \textit{lower-$Z$}-guiding layer materials shows a clear potential for improving the performance of state-of-the-art cavities and their applications.

Finally, we note that optimizing the top and the bottom cladding layer materials independently still suggests that the best performance will be achieved, if both layers are of the lowest possible absorption. 

\section{\label{sec:Isotopes}Influence of the resonant isotopes on the parameter space accessible by the inverse design}

Having discussed the accessible CLSs and SRs, the reflection spectrum visibility of the nuclear response, the field enhancement and relevant material aspects for thin-film cavities, it remains to address the influence of different resonant isotopes. Taking a key role in the properties of the artificial quantum system to be designed, their influence is two-fold. 

Firstly, each isotope comes with intrinsic properties. They determine the coupling of the nuclear transition to the electromagnetic environment and thus fix the scale of collective effects. Secondly, even ultra-thin layers of resonant isotopes modify the field configuration in the cavity by their electronic refractive index. Exchanging the resonant isotope thus strongly affects the cavity mode structure. Furthermore, the refractive indices of the cavity materials and the optimal geometries are subject to the light's wavelength and thus to the nuclear transition frequency, which thereby also influences the modal structure.

In the following, we disentangle the influence of both effects on the cavity performance and outline the consequences on the design of artificial few-level systems. To this end, we consider the accessible parameter combinations of the SR and the field enhancement at the resonant layer for the isotopes \Fe, \Sn and \Sc.  
These isotopes all have transition frequencies accessible with state-of-the-art pulsed x-ray sources, but feature rather different properties. The 
transition frequencies and electronic refractive indices of these isotopes are given in Tab. \ref{tab:ResIso}. The isotopes are chosen to be embedded in a Pd/C/isotope/C/Pd/Si cavity. We note that for better comparability, we chose the thickness of the resonant layer as $0.574\,\mathrm{nm}$ in 
all cases, as  was previously used for the \Fe cavities.  The accessible parameter combinations are shown in Fig.~\ref{fig:Pd_C_resIso_C_Pd_Si_SR_Efield_comparison.pdf}. \Sc and \Sn stand out with very high field enhancements, but comparably low tuning capability in the SR. In comparison, \Fe exhibits comparatively high possible SR in combination with moderate field enhancement. This highlights that in optimizing  x-ray cavity structures towards a given design goal, also the resonant isotope should be considered as  an important design parameter.

\begin{table}[t]
\begin{ruledtabular}
\begin{tabular}{l|c|c|c}
&$\omega_\mathrm{nuc}\,(\mathrm{keV})$ &$\delta_\mathrm{iso}+i\beta_\mathrm{iso}\,(10^{-6})$&$\delta_\mathrm{C}+i\beta_\mathrm{C}\,(10^{-6})$\\
\hline
\Fe&14.4&$7.3+0.33 i$&$2.3 +1.2\times 10^{-3} i$\\
\Sn&23.9&$2.2 +0.037 i$&$0.82 +2.8\times 10^{-4} i$\\
\Sc&12.4&$3.8+0.13i$&$3.1 +2.2\times 10^{-3} i$\\
\end{tabular}
\end{ruledtabular}
\caption{Properties of the M\"ossbauer isotopes \Fe, \Sn and \Sc and the Carbon guiding layer at the respective M\"ossbauer resonance frequencies. The electronic refractive index of the resonant isotope at its transition frequency is denoted as $n_\mathrm{isotope}=1-\delta_\mathrm{iso}+i\beta_\mathrm{iso}$ and the respective one of Carbon reads $n_\mathrm{C}=1-\delta_\mathrm{C}+i\beta_\mathrm{C}$. The isotope's transition frequency is given by $\omega_\mathrm{nuc}$. Parameters are taken 
from Ref.~\cite{rohlsberger_nuclear_2005} or calculated with the \textsc{PYNUSS}~\cite{heeg_pynuss_nodate} software package.}
\label{tab:ResIso}
\end{table}

\subsection{\label{sec:direct-nuclear}Direct impact of the nuclear properties}
To disentangle the effects of exchanging the nuclear isotope on the cavity performance, we first try to compare the results for the different isotopes via a naive scaling of their parameters. 

We start with $\CLS, \SR \propto \omega^2_{\mathrm{nuc}}|\vec{d}|^2G(\omega_\mathrm{nuc}){N}/{A}$, following Eqs.~\eqref{eq:DeltaEff} and \eqref{eq:GammaEff}. We can express the dipole moment in this relation  as~\cite{lentrodt_ab_2020}
\begin{equation}
|\vec{d}|^2=\frac{2\pi\gamma_0}{\omega_\mathrm{nuc}^3}\frac{1}{2(1+\alpha)}\frac{2I_\mathrm{e}+1}{2I_\mathrm{g}+1}\,,
\label{eq:dipoleMoment}
\end{equation}
where $\alpha$ is the coefficient of internal conversion and $I_\mathrm{e}$ ($I_\mathrm{g}$) is the nuclear spin of the excited (ground) state of the transition. Moreover, the effective in-plane nuclear number density can be written as
\begin{equation}
\frac{N}{A}=d_3\rho_\mathrm{V}af_\mathrm{LM}\,,
\end{equation}
with the Lamb-M\"ossbauer factor $f_\mathrm{LM}$, the  number density per 
volume of the material $\rho_\mathrm{N}$, the thickness of the resonant layer $d_3$ and the abundance $a$ of the resonant isotopes therein.

Regarding the Green's function~\eqref{eq:theory:5LayerGF}, we note that the Fresnel coefficients in Eq.~\eqref{eq:theory:ReflectionCoeff} are unaffected by the frequency dependence, such that we find a scaling  $G(\omega_\mathrm{nuc})\propto1/\omega_\mathrm{nuc}$ with the isotope's transition frequency, owing to $\beta_j
\propto \omega_\mathrm{nuc}$.

Taking into account all these scalings, we can estimate  the CLS and SR of a specific isotope on the basis of the numerically calculated performance of \Fe. However, upon comparing these estimates to the results in Fig.~\ref{fig:Pd_C_resIso_C_Pd_Si_SR_Efield_comparison.pdf}, we find  that the 
SRs actually calculated for \Sn and \Sc are significantly higher than those expected from the naive scaling. This is a clear indication that the nuclear transition properties alone are not sufficient to characterize the performance of an isotope in a thin-film cavity. Instead, also the modification of the cavity environment due to the exchange of the isotope layer must be considered.

\subsection{Impact of the nuclear isotope on the cavity environment}

Next, we study the influence of the isotope choice on the electromagnetic environment provided by the cavity. Exchanging the resonant isotope alters the photonic DOS as, on the one hand, each resonant isotope comes with its specific electronic refractive index, and, on the other hand, its resonance frequency affects the refractive indices of the remaining layers. As can be seen from Tab. \ref{tab:ResIso}, the carbon guiding layer's refractive index is approaching unity with increasing frequency of the isotopes, as common for x-ray materials~\cite{als-nielsen_elements_2011}. For the refractive index of the resonant isotopes themselves there is no such clear trend with respect to transition frequency as it competes with their electron density.
\begin{figure}[t]
\includegraphics[width=0.9\columnwidth]{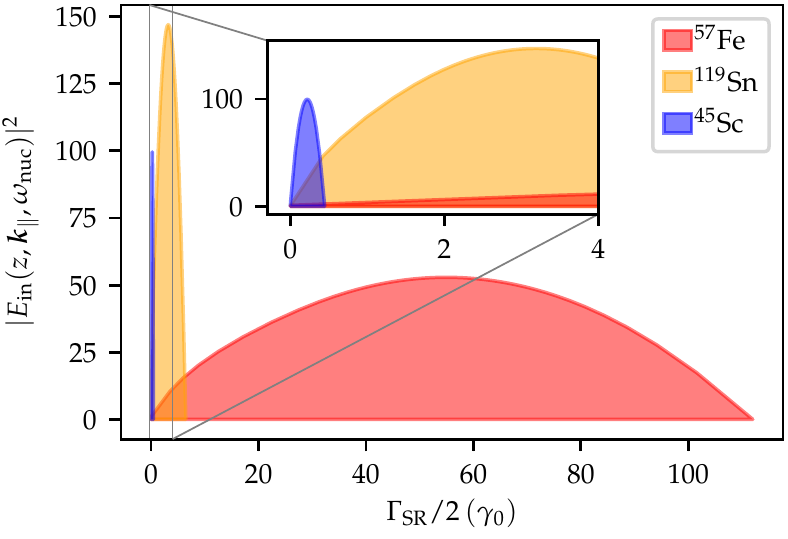}
\caption{(Color online) Accessible combinations of the SR of the artificial x-ray two-level system and  the cavity-induced field enhancement at the position of the nuclei as a function of the chosen resonant isotope. In 
all cases, the optimization was performed with a  Pd/C/isotope/C/Pd/Si cavity structure. It can be seen that cavities with \Sn and \Sc feature superior field enhancements, while the \Fe cavities offer comparably large SR in units of the respective natural line widths $\gamma_0$.}
\label{fig:Pd_C_resIso_C_Pd_Si_SR_Efield_comparison.pdf}
\end{figure}

To estimate the effect of the isotope choice on the photonic DOS, we take 
the achievable field enhancement at the nuclear layer as shown in Fig.~\ref{fig:Pd_C_resIso_C_Pd_Si_SR_Efield_comparison.pdf}. Albeit in section \ref{sec:CLS-SR-EF} we noted that field enhancement and SR are realized in 
different geometries, this is a sensible choice as we do not compare geometries but the materials of the resonant layer. Also, the results in Sec.~\ref{sec:Materials} suggest that within one setting of materials, the maximum achievable SR and field enhancement are correlated to each other. Finally, the field enhancement is particularly useful as it is irrespective of the nuclear properties.

Comparing the refractive indices in Tab. \ref{tab:ResIso} to the results in Fig.~\ref{fig:Pd_C_resIso_C_Pd_Si_SR_Efield_comparison.pdf}, it is clear  that low absorption in the cavity, and, in particular, in the resonant layer, is the driving factor for the photonic environment. The very high field enhancement for \Sn and significantly reduced enhancement for \Fe due to the electromagnetic environment complement very well the incompleteness of the description found in Sec.~\ref{sec:direct-nuclear} upon considering the nuclear properties only. 

This shows that considering the nuclear properties alone is not sufficient to estimate an isotope's potential. Rather, the effect of the isotope on the electromagnetic environment has to be taken into account for a comprehensive description.

\section{\label{sec:conclusion}Conclusion and outlook}

In summary, we introduced the inverse design of an artificial x-ray two-level system, realized by an archetype x-ray thin-film cavity with a single layer of M\"ossbauer nuclei. In particular, we investigated and explained the accessible combinations of the superradiant line broadening and the transition energy shifts of the artificial two-level system and studied their connection to the experimentally relevant visibility of the collective effects in the reflection spectra.

We further studied the interplay of the driving field enhancement at the nuclear layer with the frequency shift and decay enhancement. Strikingly, our analysis revealed that the maximal field enhancement is not found in the same cavity geometry as the largest decay enhancement. Opposed to that, we showed that a two-level system in an optical Fabry-P\'erot cavity exhibits a common optimal geometry for both quantities. We could attribute this qualitative difference to the grazing incidence nature of the x-ray cavity setting. This feature is not limited to the archetype system, discussed in this work, but can be applied to essentially all x-ray thin-film cavity systems. It thus helps to broaden the intuitive understanding of thin-film cavity systems in general which is crucial for the development of novel ideas and applications. 

To explore the possible scope of the inverse design approach, we systematically sampled different materials for guiding and cladding layers. Our results promote a new perspective on the impact of the cladding layer's absorption on the overall performance. The analysis indicates that \textit{low-Z} cladding materials outperform \textit{high-Z} ones, i.e. the (real part) refractive index contrast of guiding and cladding layer is subordinate to the cladding layer's absorption. As a result, for particular settings, cavities with no upper cladding (mirror) layer at all are found to be superior. Applying a \textit{low-Z guiding -- lower-Z cladding} design paradigm, the field enhancement in the cavity as well as collective effects can be enhanced. This insight is not only applicable to cavities beyond the archetype systems at hand, but might also prove interesting for x-ray photonic environments (e.g., one-dimendional waveguides, nano-wires, or crystal environments) in general.

Subsequently, we completed the discussion of the inverse design scope by analysing the influence of different resonant isotopes on the field enhancement and collective effects. Here, it became clear that the overall impact of the M\"ossbauer nuclei on the photonic environment of the cavity cannot be neglected for a comprehensive discussion. 

Our results can readily be extended to more general artificial x-ray level schemes realized with x-ray thin-film cavities featuring multiple isotope layers. They therefore set the stage for the future inverse design of more advanced and intriguing quantum optical schemes, thus furthering the field of x-ray quantum optics. 

\begin{acknowledgments}
 OD gratefully acknowledges support by the Cusanuswerk and the Studienstiftung des deutschen Volkes. 
\end{acknowledgments}

\appendix
\section{Derivation of the effective Hamiltonian in the low-excitation regime}
\label{sec:appendix:GreenFormalism}
Following Ref.~\cite{lentrodt_ab_2020}, here, we revise the derivation of the effective two-level scheme in the low-excitation regime. Starting from the many-body description (\ref{eq:Master}--\ref{eq:many2}) the equation of motion for the expectation value $\sigma_n^-\equiv\expval{\opsigma_n^-}$ in the low-excitation regime $\sigma_n^z\approx -1$ reads
\begin{align}
\dot{\sigma}_n^- = &-i\left(\omega_\mathrm{nuc}-i\frac{\gamma_0}{2}\right)\sigma_n^-\notag\\&+i\sum\limits_{n'} \left(J_{nn'}+i\frac{\Gamma_{nn'}}{2}\right)\sigma_{n'}^-+i\vec{d}^*\cdot\vec{E}(\vec{r}_n)\,.
\end{align}
Inserting this equation into the definition~\eqref{eq:SpinWaveOps} and approximating the nuclear layer to be homogeneous, the discrete Fourier transforms become continuous ones and we can rewrite the equation of motion for the expectation value of the spin-wave operator $\sigma^-_{\vec{k}_\parallel}\equiv \langle\opsigma^-_{\vec{k}_\parallel}\rangle$ as
\begin{align}
\dot{\sigma}^-_{\vec{k}_\parallel} = &-i(\omega_\mathrm{nuc}+\Delta_\mathrm{CLS}-i\frac{\gamma}{2}-i\frac{\Gamma_\mathrm{SR}}{2})\sigma^-_{\vec{k}_\parallel}\notag\\
&+ i\frac{N}{A}\vec{d}^*\cdot\vec{E}_{\mathrm{in}}(z, \vec{k}_\parallel)\,.
\end{align}

Clearly, a single spin-wave is only coupled to itself and the dynamics bound to the respective subspace. 
The very same equation is retrieved upon calculating the dynamics of $\sigma^-_{\vec{k}_\parallel}$ from the effective description~(\ref{eq:single1},~\ref{eq:single2}) when assuming linear response, $\sigma^z_{\vec{k}_\parallel}\approx -1$. Hence, in this regime, the dynamics is equally well characterized by the effective few-level description. 

Fourier transforming $\sigma^-_{\vec{k}_\parallel}(\omega)=\int\dd{t}e^{-i\omega t}\sigma^-_{\vec{k}_\parallel}$, the frequency space solution is readily found to be 
\begin{equation}
\sigma^-_{\vec{k}_\parallel}(\omega) = -\frac{\vec{d}^*\cdot\vec{E}_\mathrm{in}(z, \vec{k}_\parallel, \omega_\mathrm{nuc})}{\omega-\omega_\mathrm{nuc}-\Delta_\mathrm{CLS}+i({\gamma_0}+{\Gamma_\mathrm{SR}})/2}\,.
\label{eq:frequencySpaceSolution}
\end{equation}

The collective dynamics of the nuclei at position $z$ modify the overall electric field $\vec{E}(0, \vec{k}_\parallel, \omega)$ at the surface which can be calculated by the generalized input-output relation~\cite{asenjo-garcia_atom-light_2017}
\begin{align}
\vec{E}(0, \vec{k}_\parallel, \omega)=&\vec{E}_\mathrm{in}(0, \vec{k}_\parallel, \omega_\mathrm{nuc})\notag\\&+\mu_0\omega_\mathrm{nuc}^2\vec{G}(0, z, \vec{k}_\parallel, \omega_\mathrm{nuc})\cdot\vec{d}\, \sigma^-_{\vec{k}_\parallel}(\omega)\,.
\label{eq:theory:InputOutput}
\end{align}
Note that in Eqs.~(\ref{eq:frequencySpaceSolution}) and (\ref{eq:theory:InputOutput}) we used the fact that the incoming electric field and the Green's function can be approximated as constant in frequency on scales of the nuclear linewidth.

Finally, for the incoming field strength normalized to one, the overall reflection coefficient is given by subtracting the incident field strength from the overall electric field (here without polarization), 
\begin{align}
r(\vec{k}_\parallel, \omega) &= {{E}(0, \vec{k}_\parallel, \omega)}-1 \notag\\
&=r_\mathrm{el} + \mu_0\omega_\mathrm{nuc}^2\, G(0, z, \vec{k}_\parallel, \omega)\,d \, \sigma^-_{\vec{k}_\parallel}(\omega)\,,
\label{eq:theory:reflection}
\end{align}
where we used the electronic cavity reflectivity 
\begin{align}
&r_\mathrm{el} = {E}_\mathrm{in}(0, \vec{k}_\parallel, \omega_\mathrm{nuc})-1\,,
\label{eq:theory:paramsReflR}
\end{align}
given by the (cavity modified) electric field at the cavity surface without the incoming field strength. This, in turn, yields Eq.~\eqref{eq:theory:Fano} in the main text.

\section{Relevant Green's function evaluations and field configurations for the archetype cavity}\label{sec:appendix:formulae}
Here, we summarize the formulae for the Green's function and field configuration as needed for the explicit calculation of the observables in the archetype cavity of Fig.~\ref{fig:schematicSetup}. The formulae are taken from Ref.~\cite{tomas_green_1995}.

\subsection{Green's function}\label{sec:appendix:GFs}
The in-plane Fourier transformed Green's function for s-polarization at equal $z$ in the third (resonant) layer is given by
\begin{align}
&{G}(z, z, \vec{k}_\parallel, \omega) = \frac{i}{2\beta_3}\frac{e^{i\beta_3 d_3}}{1-r_{3/0}r_{3/6}e^{2i\beta_3d_3}}\notag\\
&\quad\times(e^{i\beta_3(z-d_3)}+r_{3/6}e^{-i\beta_3(z-d_3)})(e^{-i\beta_3z}+r_{3/0}e^{i\beta_3z})\,,
\label{eq:theory:5LayerGF}
\end{align}
where
\begin{align}
\begin{aligned}[c]
&r_{3/0}=\frac{-r_{23}+r_{2/0}e^{2i\beta_2d_2}}{1-r_{23}r_{2/0}e^{2i\beta_2d_2}}\,,\\
&r_{3/6}=\frac{r_{34}+r_{4/6}e^{2i\beta_4d_4}}{1+r_{34}r_{4/6}e^{2i\beta_4d_4}}\,,
\end{aligned}
\quad
\begin{aligned}[c]
&r_{2/0}=-\frac{r_{12}+r_{01}e^{2i\beta_1d_1}}{1+r_{12}r_{01}e^{2i\beta_1d_1}}\,,\\
&r_{4/6}=\frac{r_{45}+r_{56}e^{2i\beta_5d_5}}{1+r_{45}r_{56}e^{2i\beta_5d_5}}\,.
\end{aligned}
\label{eq:theory:ReflectionCoeff}
\end{align}
$r_{ij}$ denotes the Fresnel coefficient of light in layer $i$ reflected  at adjacent layer $j$. Furthermore, $\beta_j = \sqrt{k_j^2-\vec{k}_{\parallel}^2}$, where $k_j=n_j\omega$ are the wavenumbers in layer $j$ obtained from the refractive indices $n_j$. The thicknesses $d_j$ are enumerated according to Fig.~\ref{fig:schematicSetup} and $z$ denotes the 
distance to the third layer top surface which we will generally set to the center of the ultrathin resonant layer $z=d_3/2$. It is noted, that simpler cavity structures can be obtained by setting the respective thicknesses to zero. The additional $\delta$-contribution to the Green's function, apparent in~\cite{tomas_green_1995}, renormalizes the free-space transition frequency and decay. It is thus taken to be already included in the respective parameters~\cite{lentrodt_ab_2020}.

The Green's function propagating the nuclear response to the surface is further given by 
\begin{align}
{G}(0, z, \vec{k}_\parallel; \omega)= &\frac{i}{2\beta_0}\frac{t_{0/3}e^{i\beta_3d_3}}{1-r_{3/0}r_{3/6}e^{2i\beta_3d_3}}\notag\\
&\times\left(e^{-i\beta_3(d_3-z)}+r_{3/6}e^{i\beta_3(d_3-z)}\right)\,.
\end{align} 
The additional coefficients are defined as
\begin{equation}
\begin{aligned}
t_{0/3} = \frac{t_{0/2}t_{23}e^{i\beta_2d_2}}{1-r_{2/0}r_{23}e^{2i\beta_2d_2}}\,,
\end{aligned}\quad
\begin{aligned}
t_{0/2} = \frac{t_{01}t_{12}e^{i\beta_1d_1}}{1+r_{01}r_{12}e^{2i\beta_1d_1}}\,,
\end{aligned}
\end{equation}
for $t_{ij}$ being the Fresnel coefficients of light in layer $i$ being transmitted to adjacent layer $j$.

\subsection{Field configuration}\label{sec:appendix:fields}
The electric field strength at the third, resonant layer is given as~\cite{tomas_green_1995}
\begin{align}
{E}_{\mathrm{in}}(z, \vec{k}_\parallel, \omega) &= \frac{t_{0/3}e^{i\beta_3d_3}}{1-r_{3/0}r_{3/6}e^{i\beta_3d_3}}\notag\\ &\quad\times\left(e^{i\beta_3(z-d_3)}+r_{3/6}e^{-i\beta_3(z-d_3)}\right)\,,
\label{eq:appendix:fieldNuclei}
\end{align}
where as before we evaluate the field at the center of the ultrathin layer, $z=d_3/2$. 

For the calculation of the electronic cavity reflection the field strength at the surface is used,
\begin{align}
&{E}_\mathrm{in}(0, \vec{k}_\parallel, \omega) = 1+r_\mathrm{el} = 1+\frac{1}{1-r_{3/0}r_{3/6}e^{2i\beta_3d_3}}\notag\\
&\qquad\times\left[{r_{0/3}+(t_{0/3}t_{3/0}-r_{0/3}r_{3/0})r_{3/6}e^{2i\beta_3d_3}}\right]\,,
\end{align}
where the additional coefficients are
\begin{equation}
\begin{aligned}
r_{0/3} = \frac{r_{01}+r_{1/3}e^{2i\beta_1d_1}}{1+r_{01}r_{1/3}e^{2i\beta_1d_1}}\,,\\
t_{3/0}=\frac{t_{3/1}t_{10}e^{i\beta_1d_1}}{1+r_{1/3}r_{01}e^{2i\beta_1d_1}}\,,
\end{aligned}
\quad
\begin{aligned}
r_{1/3}=\frac{r_{12}+r_{23}e^{2i\beta_2d_2}}{1+r_{12}r_{23}e^{2i\beta_2d_2}}\,,\\
t_{3/1}=\frac{t_{32}t_{21}e^{i\beta_2d_2}}{1+r_{23}r_{12}e^{2i\beta_2d_2}}\,.
\end{aligned}
\end{equation}

\section{Numerical methods}\label{sec:appendix:numericalMethods}
We use the \texttt{scipy.optimize} package for the determination of the parameter spaces throughout this work. The surfaces of these multidimensional sets, however, are not directly accessible by scalar optimization. Rather, one has to find a suitable scalar cost function of the observables which can then be passed on to the optimization algorithm.

\begin{figure}[t]
    \includegraphics[]{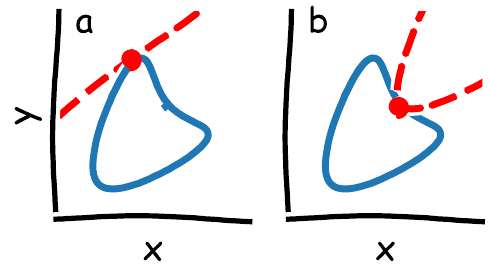}
    \caption{(Color online) Optimization methods applied to a non-convex set. Red lines represent constant cost function for a linear (a) and nonlinear (b) combination of observables.}
    \label{fig:SetBoundaryFinding_schematic.pdf}
\end{figure}

The most straightforward way of doing this is taking linear combinations of the observables. For an exemplary, two-dimensional set in Fig.~\ref{fig:SetBoundaryFinding_schematic.pdf}(a), the cost function $f = \alpha x + \beta y$ is constant on some line in the plane. Correspondingly, a converged maximization will yield the intersection point of the furthest tangent $f'=\alpha x + \beta y +\mathcal{C}$ and the set. By changing $\alpha$ and $\beta$, the different slopes of the line will be achieved. However, this procedure may only work, if the underlying set is convex -- which is not always the case for the above parameter spaces.

To overcome this problem, one can find a suitable non-linear combination of observables when keeping in mind the manifold where this function is supposed to be constant. For a star-convex set, for example, we can use a narrow, adequately placed parabola to uniquely determine the set boundary, as indicated in Fig.~\ref{fig:SetBoundaryFinding_schematic.pdf}(b). To retrieve the corresponding scalar cost function $f$, we consider the parabola being an upright parabola transformed by rotation matrix $\vec{R}$, 

\begin{align}
&\vec{R}\left\lbrace(x, y)\in\mathbb{R}^2|y-x^2=\mathcal{C}\right\rbrace\notag \\ &= \left\lbrace(x, y)\in\mathbb{R}^2|y'-x'^2=\mathcal{C},
\left(\begin{matrix}
x'\\
y'
\end{matrix}\right)
=\vec{R}^{-1}
\left(\begin{matrix}
x\\
y
\end{matrix}\right)
\right\rbrace\,.
\end{align}

From the second, equivalent, expression we can then directly read of the cost function to be $f=y'-x'^2$, with the primed coordinates replaced by the linear combination of $x$ and $y$. The constant $\mathcal{C}$ is of no interest, as it only changes the value of the cost function but not the position where the optimum is achieved.

\end{document}